\documentclass{aastex63}

\usepackage[caption=false]{subfig}

\newcommand{\eqref}{eq(\ref}
\newcommand{\sag}{Sagittarius}
\newcommand{\gaia}{\textit{Gaia}}
\newcommand{\msat}{$M_{sat}$}
\newcommand{\TPdf}{\texttt{3D quasiisothermaldf}}
\newcommand{\plummerpot}{\texttt{PlummerPotential}}
\newcommand{\chandrasekhardff}{\texttt{ChandrasekharDynamicalFrictionForce}}
\newcommand{\movingobjpot}{\texttt{MovingObjectPotential}}
\newcommand{\Delvz}{\ensuremath{\Delta v_z}}
\newcommand{\tenexp}{\times 10^}
\newcommand{\msun}{\ensuremath{M_\odot}}
\newcommand{\az}{\ensuremath{a_z}}
\newcommand{\Msg}{\ensuremath{M_{\text{Sgr}}}}
\newcommand{\phisat}{\ensuremath{\phi_{\text{sat}}}}
\newcommand{\phistar}{\ensuremath{\phi_{\star}}}
\newcommand{\vstar}{$v_{\star}$}
\newcommand{\delphi}{\phisat-\phistar(\timp)}

\newcommand{\jphi}{\ensuremath{J_\phi}}
\newcommand{\galpy}{\texttt{galpy}}
\newcommand{\galpys}{\texttt{galpy's}}
\newcommand{\mwpot}{\texttt{MWPotential2014}}
\newcommand{\timp}{\ensuremath{t_{\text{imp}}}}
\newcommand{\azvz}{$A_{z,v_z}$}
\newcommand{\omegaz}{$\Omega_z$}
\newcommand{\omegaphi}{$\Omega_\phi$}
\newcommand{\xy}{$x$--$y$}

\newcommand{\kmpers}{km$s^{-1}$}

\newcommand{\jphivo}{\ensuremath{J_\phi/v_0}}
\newcommand{\figref}{Figure~\ref}
\newcommand{\figsref}{Figures~\ref}
\newcommand{\rgal}{\ensuremath{R}}
\newcommand{\z}{\ensuremath{z}}
\newcommand{\rg}{\ensuremath{R_g}}

\newcommand{\vz}{\ensuremath{v_z}}

\newcommand{\rphi}{$R$--$\phi$}

\newcommand{\zvz}{$z$--$v_z$}

\received{---}
\revised{---}
\submitjournal{The Astrophysical Journal}

\shorttitle{Snails Across Scales}
\shortauthors{S. S. Gandhi et al.}

\begin{document}

\title{Snails Across Scales: \\  Local and Global Phase-Mixing Structures as Probes of the Past and  Future  Milky Way.}

\correspondingauthor{Suroor S. Gandhi}
\email{ssg487@nyu.edu}

\author[0000-0001-5640-8636]{Suroor S. Gandhi}
\affiliation{Center for Cosmology and Particle Physics, New York University, 726 Broadway, New York, NY 10003, USA}

\author{Kathryn V. Johnston }
\affiliation{Center for Computational Astrophysics, Flatiron Institute, 162 5th Ave., New York City, NY 10010, USA}
\affiliation{Department of Astronomy, Columbia University, New York, NY 10027, USA}


\author{Jason A.\ S.\ Hunt }
\affiliation{Center for Computational Astrophysics, Flatiron Institute, 162 5th Ave., New York City, NY 10010, USA}

\author{Adrian M. Price-Whelan}
\affiliation{Center for Computational Astrophysics, Flatiron Institute, 162 5th Ave., New York City, NY 10010, USA}

\author[0000-0003-3922-7336]{Chervin F. P. Laporte}
\affiliation{Institut de Ci\`encies del Cosmos (ICCUB), Universitat de Barcelona (IEEC-UB), Mart\'i i Franqu\`es 1, 08028 Barcelona, Spain}
\affiliation{Kavli Institute for the Physics and Mathematics of the Universe (WPI), The University of Tokyo Institutes for Advanced Study (UTIAS), \hspace{0.22cm}The University of Tokyo, Chiba 277-8583, Japan}

\author{David W. Hogg}
\affiliation{Center for Cosmology and Particle Physics, New York University, 726 Broadway, New York, NY 10003, USA}
\affiliation{Center for Computational Astrophysics, Flatiron Institute, 162 5th Ave., New York City, NY 10010, USA}
\affiliation{Max-Planck-Institut f\"{u}r Astronomie, K\"{o}nigstuhl 17, D-69117 Heidelberg, Germany}

\begin{abstract}
    Signatures of vertical disequilibrium have been observed across the Milky Way's disk. These signatures manifest locally as unmixed phase-spirals in \zvz\ space (``snails-in-phase”) and globally as nonzero mean $z$ and $v_z$ which wraps around as a physical spiral across the \xy-plane (``snails-in-space”). We explore the connection between these local and global spirals through the example of a satellite perturbing a test-particle Milky Way (MW)-like disk. We anticipate our results to broadly apply to any vertical perturbation.
    
Using a \zvz-asymmetry metric we demonstrate that in test-particle simulations: (a) multiple local phase-spiral morphologies appear when stars are binned by azimuthal action \jphi, excited by a single event (in our case, a satellite disk-crossing); (b) these distinct phase-spirals are traced back to distinct disk locations; and (c) they are excited at distinct times. Thus, local phase-spirals offer a global view of the MW’s perturbation history from multiple perspectives. 

Using a toy model for a Sagittarius (Sgr)-like satellite crossing the disk, we show that the full interaction takes place on timescales comparable to orbital periods of disk stars within $R\lesssim10$ kpc. Hence such perturbations have widespread influence which peaks in distinct regions of the disk at different times.

This leads us to examine the ongoing MW-Sgr interaction. While Sgr has not yet crossed the disk (currently, $z_{\mathrm{Sgr}}\approx -6$ kpc, $v_{z,Sgr}\approx210$ \kmpers), we demonstrate that the peak of the impact has already passed. Sgr’s pull over the past 150 Myr creates a global \vz\ signature with amplitude $\propto \Msg$, which might be detectable in future spectroscopic surveys.

\end{abstract}



\section{Introduction} \label{sec:intro}

Our understanding of vertical structure in the Milky Way has drastically evolved over the past decade. Discussions of the disk have traditionally been centered around equilibrium axisymmetric models ---  planar and fully phase-mixed vertically and with simple periodic perturbations in azimuth. The emerging field of galactoseismology is shifting that focus towards the mild, but significant departures from equilibrium that are increasingly evident in observations. Vertical asymmetries were first pointed out in three distinct data sets by \citet{widrow2012discovery, carlin2013}, and \citet{ williams2013}. \cite{widrow2012discovery} found North-South asymmetry in SDSS DR-8 \citep{Aihara_sdss_dr8_2011} and SEGUE \citep{Yanny_segue_2009}. \cite{carlin2013} used PPMXL \citep{roeser2010ppmxl} and LAMOST \citep{cui2012lamost, zhao2012lamost} to conclude that stars above and below the midplane exhibit opposite radial motion. \citet{williams2013} found similar asymmetry around the solar neighborhood in RAVE \citep{steinmetz2006rave}.

These local asymmetries in disk motions were shown to be matched by vertical asymmetries of the disk in space, which were traced to several  kpc beyond the Sun by \citet{xu2015}. Coincidentally, \citet{price_whelan2015} were finding evidence that structures tens of degrees from the  plane and at Galactocentric radii of 15-30 kpc, well beyond the traditional limits of the disk \citep[see][for discovery papers]{Newberg02}  nevertheless had velocity trends and stellar population properties consistent with disk membership \citep[see also subsequent work that confirms this interpretation][]{Sheffield18,Li17,Bergemann18}. These discoveries suggested the local corrugations of the disk were likely part of a global pattern of bending and breathing modes, as first pointed out by \cite{widrow_barber2014,gomez2013}.
Simulations of LMC and Sgr-like satellites interacting with a Milky-Way-scale galaxy could reproduce the scales of these perturbations, both locally and globally, supporting the plausibility of this interpretation of local and global-scale asymmetries being associated  \citep{laporte2018b, laporte2018a}.


The reach and high-dimensionality of the \gaia\ data sets \citep{gaia_dr1_2016, gaia_dr2_2018,gaia_edr3_2020} allowed clear confirmation of what these earlier studies were hinting at --- the existence of global-scale, vertical ripples coursing through our Galactic disk \citep{gaia_dr2_disk_paper}.
For the first time, \gaia\ DR-2 enabled the local vertical asymmetries in position and velocity to be dramatically visualized as a clear \zvz\ phase-spiral\footnote{\footnotesize{ We refer to the spiral structures as `snails' occasionally. Terms most commonly used for the spirals in a \textit{local} volume are `\zvz\ spiral', `phase-space spiral', or `phase-spiral.' We will mostly use the term `phase-spiral.' Spirals that form \textit{globally} across the \xy\ plane of the disk are referred to as `\rphi\ spirals' or `physical spirals' in this work.} } \citep{antoja2018}. 
The richness of the data have inspired  analysis and comparison to simulations on both global \citep[e.g. see projections and visualizations in][]{schonrich_dehnen2018,kawata2018,Salomon2020,poggio2018a,poggio2018b,laporte2019footprints,poggio2020,Eilers+20,FS19} and local \citep{antoja2018,binney_schonrich2018,darling_widrow2019,laporte2019footprints,bland-hawthorn2019,li2020} scales.
In particular, simulations of a Sgr-like satellite impacting a MW-like disk that were developed to fit the pre-\gaia\ data were shown to contain analogous manifestation of both the local and global signatures of vertical disequilibrium with similar spatial and velocity scales \citep{laporte2019footprints}. Other works have reinforced these local-global connections with further views of the data, tailored simulations and analytic models \citep{xu_liu_tian2020,bland-hawthorn_tepper-garcia2020,widrow_darling_li2020,bennett_bovy2021}. 

While the impact of a satellite provides a natural explanation for the origin of oscillations perpendicular to the Galactic Plane, it is not the only one. Any vertical perturbation, such as bar buckling \citep{khoperskov2019}, can plausibly do the same \citep[although stellar ages of the phase-spiral suggest this is not the origin in the Milky Way, see][]{laporte2019footprints} . Moreover, once  bending motions are excited in the disk, self-gravity can launch further disturbances \citep{darling_widrow2019,bland-hawthorn_tepper-garcia2020}.

Luckily, the combination of local and global responses should provide multiple constraints on the origin of specific features.
The coupled evolution of these spirals is largely driven by the simple process known as {\it phase-mixing} \citep[see][for some limitations of this interpretation]{darling_widrow2019}. Phase mixing can occur when stars which are initially at \textit{random} orbital phases are systematically offset by a perturbation to then be at the \textit{same} orbital phase.
If these stars have a range of orbital properties (e.g. frequencies) they will subsequently spread along the orbit and {\it mix} in orbital phase. \figref{fig:freqs_and_time_periods} shows the different orbital frequencies as a function of guiding radius, $R_g$ (a proxy for distance from the Galactic center; see \S\ref{subsec: dissect-snail} for definition). In the disk \xy\ plane phase-mixing is driven by the $R_g$-dependence of azimuthal frequencies,  $\Omega_\phi$, while in 
the \zvz-plane, phase-mixing is driven by the range in vertical frequencies, $\Omega_z$, at any fixed $R_g$ (see \figref{fig:freqs_and_time_periods}b). 
Phase-mixing can lead to a well-defined spiral for some time, but ultimately this spiral will wind up to the extent that the population once again appears to have randomly distributed phases.

Conceptually, rewinding and interpreting the \zvz\ and \rphi\ spirals \textit{simultaneously} should give us insight into the timing, strength and location of the perturbation that caused them, adding clarity to our understanding of the nature of that perturbation as well as the properties of orbits across the Galactic disk.
This paper explores the feasibility of this ultimate goal. Our study uses both test particle and N-body simulations (described in \S 2) to explore how multiple signatures can be traced back to offer multiple views of a single event (described in \S 3). The results are used in a first application of the ongoing interaction between Sgr and the MW (described in \S 4) and future prospects are discussed in \S 5.

\section{Simulations \& Data} \label{sec: simulations}
In this section we describe the simulations and data that we analyze in subsequent sections. Although this paper is based on \gaia\ eDR-3 data, we interpret those data in terms of simple simulations. After presenting relevant observations in \gaia\ data, we primarily use test particle simulations which allow us to fully control the orbit of the satellite and the galactic potential. These are described in \S\ref{subsec: tp sim description}. We also show that our results are robust in the presence of self gravity by including a comparison to a self-consistent simulation (in \S\ref{sec: results-III_ongoing_interaction}), and the self-consistent model is described in \S\ref{subsec: bonsai sim description}.

\begin{figure}
    \centering
    \includegraphics[width=\textwidth]{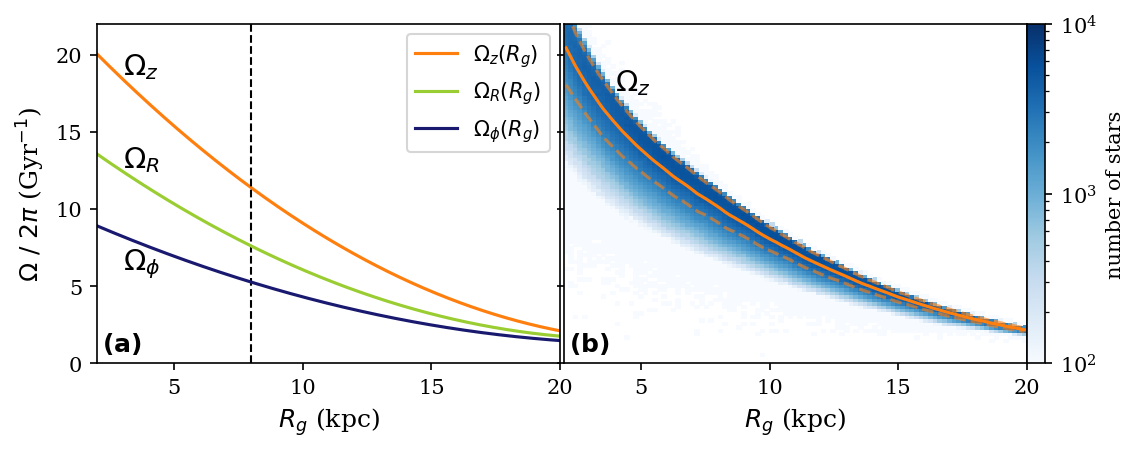}
    \caption{\textbf{(a)} Best-fit curves for epicyclic oscillation frequencies expressed as $\Omega/2\pi$(Gyr$^{-1}$) for simulated stars in an unperturbed disk in the $z$, $R$, and $\phi$ directions as a function of \rg=\jphivo. The $\Omega(\rg)$ curves in this figure are fit to near-circular orbits, i.e., for which $\sqrt{J_R^2 + J_z^2}/J_\phi \ll 1$, where $J_R$ and $J_z$ are orbital actions in the \rgal\ and $z$ directions respectively. \textbf{(b)} A 2D histogram showing the spread in $\Omega_z$ as a function of \rg\, not restricted to near-circular orbits anymore. The solid orange curve in this panel (different from the orange curve in panel (a)) is the median \omegaz, and the dashed curves mark the $1\sigma$ dispersion about the median. The spread in $\Omega_z$ \textit{at a particular} \rg\ value, $R_{g\star}$, is what causes a phase-spiral to develop in the $z$-$v_z$ plane. Stars at the head of a \zvz~spiral have high $\Omega_z(R_{g\star})$, whereas stars in the spiral tail have lower $\Omega_z(R_{g\star})$. The \textit{variation of \omegaz\ with} \rg\ is what causes \textit{multiple} spiral morphologies to develop once the disk is perturbed.}
    \label{fig:freqs_and_time_periods}
\end{figure}

All the simulated MW-like disks that we use in this work generically exhibit  orbital properties illustrated in  \figref{fig:freqs_and_time_periods}. \figref{fig:freqs_and_time_periods}a shows how $\phi$-rotation frequency \omegaphi, vertical epicyclic frequency \omegaz, and radial epicyclic frequency $\Omega_R$ vary with guiding radius, \rg\ (a proxy for distance from the Galactic center; see \S\ref{subsec: dissect-snail} for definition), in an unperturbed simulated disk with stars on near-circular orbits. 

\subsection{Test Particle Simulations}\label{subsec: tp sim description}
The test particle models are constructed and evolved as described in \citet{hunt2019signatures}, using the galactic dynamics library \texttt{galpy} \citep{Bovy2015}. The initial condition for the disc of massless particles is sampled from \galpys~\TPdf~ distribution function \citep[adapted from][]{binney2010}. The distribution function has an initial scale radius $R_s=R_0/3$, local radial velocity dispersion $\sigma_{v_R}= 0.15v_c(R_0)$, and local vertical velocity dispersion $\sigma_{v_z}=0.075v_c(R_0)$, where $R_0 = 8$  kpc and $v_c(R_0) = 220$ km s$^{-1}$. We evolve the disk in \texttt{galpy's} \mwpot\ for 7  Gyr allowing it to reach equilibrium.

We present three test particle models with varying parameters for the satellite galaxy as summarised in Table \ref{TPparam}. In each model, the satellite galaxy is created with \galpys~\plummerpot~as a Plummer sphere of mass $M_{sat}$ and scale parameter of 0.8. We calculate the orbit of the satellite by evolving its `present day' coordinates backwards in \mwpot~while using the \chandrasekhardff routine to take into account dynamical friction. The satellite potential then follows this orbit forward in time with \movingobjpot. The three models differ in: (i) values of \msat, (ii) initial phase-space coordinates for the satellite, (iii) time since present day for which the disk is evolved, and (iv) number of test particles in the disk.  We do not include a bar or spiral arms in the model potentials, such that the satellite galaxy is the only perturbing influence on the disk.

\hyperref[tab:I]{Model A} consists of a 200 million particle disk interacting with a satellite of mass \msat$=2\tenexp{10}\msun$ initialized with Sagittarius-like phase-space coordinates from Simbad\footnote{http://simbad.u-strasbg.fr/simbad/sim-id?Ident=sdg} \citep{simbad, simbad_coords2004,simbad_pm1997,simbad_rv2012} using \galpys\ \texttt{orbit.from\_name} routine. We use this model in \S\ref{subsec: tracing_back_spirals}. Although the satellite is the sole perturber to the disk, our conclusions about vertical disequilibrium signatures in \S\ref{subsec: results_I_conc} will apply broadly to generic perturbations (e.g. bar buckling, spiral arms, etc). The mass of the satellite is chosen to be heavier than the remnant mass of Sagittarius \citep[e.g.][]{Vasiliev2020} in order to generate a strong response, and it is held constant throughout to simplify our model. 

\hyperref[tab:I]{Models B} and \hyperref[tab:I]{C} are used in \S\ref{subsec: current_interaction_sim_prediction}, and consist of a 1 billion test particle disk. The orbit of the satellite in both models is initialized with Sagittarius' present day phase-space coordinates which we set to be $(x, y, z) = (17.5,\ 2.5,\ -6.5)$  kpc, and  $(v_x, v_y, v_z)=\ 237.9,\ -24.3,\ 209.0$) km s$^{-1}$, following \cite{Vasiliev2020}. Model B has a satellite of mass \msat$=3\tenexp{9}\msun$ and the disk has evolved under the influence of the satellite for 150 Myr (ending at the present day), such that the disk experiences only the final passage of \sag. Model C has \msat$=10^{10}\msun$ and the disk has been evolved for 4  Gyr (ending at the present day).

\begin{center}
\begin{table}[h]  \label{tab:I}
\caption{The different models used for test particle simulations (see \S\ref{subsec: tp sim description} for discussion)}

\begin{tabular}{ c|ccc } 
 \toprule
 &  \textbf{Model A}&  \textbf{Model B} & \textbf{Model C} \\ 
 \hline
 \msat & $2\tenexp{10}\msun$ & $3\tenexp{9}\msun$ & $10^{10}\msun$ \\ 
 
 Satellite ICs & Simbad & \cite{Vasiliev2020} & \cite{Vasiliev2020}  \\ 
 
 \#disk stars & 2$\tenexp8$ & $10^9$ & $10^9$  \\ 
 Disk evolution over the past & 4  Gyr & 150 Myr & 4  Gyr \\
 used in \S&\ref{subsec: tracing_back_spirals}  & \ref{subsec: current_interaction_sim_prediction} & \ref{subsec: current_interaction_sim_prediction} \\ 
 \hline
\end{tabular}
\label{TPparam}
\end{table}
\end{center}

\subsection{Self-Consistent Simulations}\label{subsec: bonsai sim description}
We also present a self-consistent model for a qualitative comparison to the test particle Models B \& C in \S\ref{subsec: current_interaction_sim_prediction}. The initial conditions are generated using \texttt{galactics} \citep{kuijken1995}, using the parameters of Model MWb from Table 2 of \cite{widrow2005} which result in a disk which remains stable against bar formation over a period of several  Gyr \citep[see][for a thorough analysis of the isolated MW-like disk galaxy]{widrow2005}. The model contains $\sim1.12\times10^9$ self gravitating particles, of which $\sim2.2\times10^8$ are in the disk, $\sim2.2\times10^7$ are in the bulge and $\sim8.8\times10^8$ are in the dark halo. 

The initial conditions for the satellite are the same as those for Sagittarius in the L2 model of \citet{laporte2018a}, which is composed of two Hernquist spheres \citep{H90}. The first represents the dark matter with $M_{200}=6\times10^{10}$ M$_{\odot}$, $c_{200}=28$, $M_h=8\times10^{10}$ M$_{\odot}$ and $a_h=8$  kpc, and the second the stellar component with $M_*=6.4\times10^8$ M$_{\odot}$ and $a_h=0.85$  kpc.

The combined model is evolved using the \texttt{Bonsai} $N$-body tree code \citep{bedorf2012bonsai} for 8.3  Gyr with a smoothing length of 50 pc and an opening angle $\theta_0=0.4$ radians. The ``present day" snapshot in \figref{fig:xy_current_interaction_present_day_different_models}c is chosen based on when the satellite is closest to the current coordinates of \sag~\citep{Vasiliev2020} with respect to the Sun, which happens to be at $t=6.88$ Gyr. The model will be released alongside a more detailed analysis in \cite{hunt2021}.

\subsection{Data}

We select stars from \gaia\ eDR-3 for which 6-D phase space information (parallax, line-of-sight velocity, sky positions, and proper motions) is available. Following \cite{antoja2018}, we require that parallax $\omega$ be positive, and that parallax error $\sigma_\omega$ be less than 20\% ($\sigma_\omega/\omega < 0.2$). We use parallax as a proxy for distance ($d=1/\omega)$, and we limit our sample to stars within $7\leq R$(kpc)$\leq 9$. This selection contains $\sim4.6$ million stars.

\section{Results I: Spirals Across Local and Global Scales} \label{sec: results-I}

Our aim is to explore the origin and evolution of \zvz\ spirals and understand how these local features relate to the macroscopic vertical ripples in the disk \xy\ plane. 
There are three factors to consider in the response to a disk perturbation: \textbf{(i)} phase-mixing around the disk in \xy\  following the perturbation creating \rphi\ spirals (\S \ref{subsubsec: x-y phase mixing}), \textbf{(ii)} phase-mixing in \zvz\ following the perturbation to form phase-spirals (\S \ref{subsubsec:z-vz phase mixing}), and \textbf{(iii)} the self-consistent disk response causing additional effects. We make the deliberate choice to focus on the combination of the first two phenomena --- phase-mixing across dimensions --- and defer the addition of the third to future work \citep[see][for some cautionary notes on the limitations of our work]{darling_widrow2019}.
In the test particle simulations, the only perturbation which can cause the onset of spirals is the satellite galaxy crossing the disk. This allows us to isolate the time and spatial scales of phase-mixing without the confusion of multiple sources of perturbations. Hence, our conclusions will apply to phase-mixing following any generic vertical perturbation to the disk, but we are missing the effect of self-gravity which can complicate the picture.

We dissect a local sample of \gaia\ data and an analogous one in the test particle simulations and uncover multiple phase-spirals within each selection (\S \ref{subsec: dissect-snail}). We explore the global context of these phase-spirals in two ways-- first by following their evolution backwards through time in the test particle simulations (\S \ref{subsec: tracing_back_spirals}) to understand when and where they were excited, and subsequently by building a toy model to illustrate the spatial and time-scales of the interaction that excited them (\S \ref{subsec: toy_model}). We put together our findings in a combined picture of phase- and physical spirals in \S \ref{subsec: results_I_conc}.

\begin{figure}
    \centering
    \includegraphics[width=\textwidth]{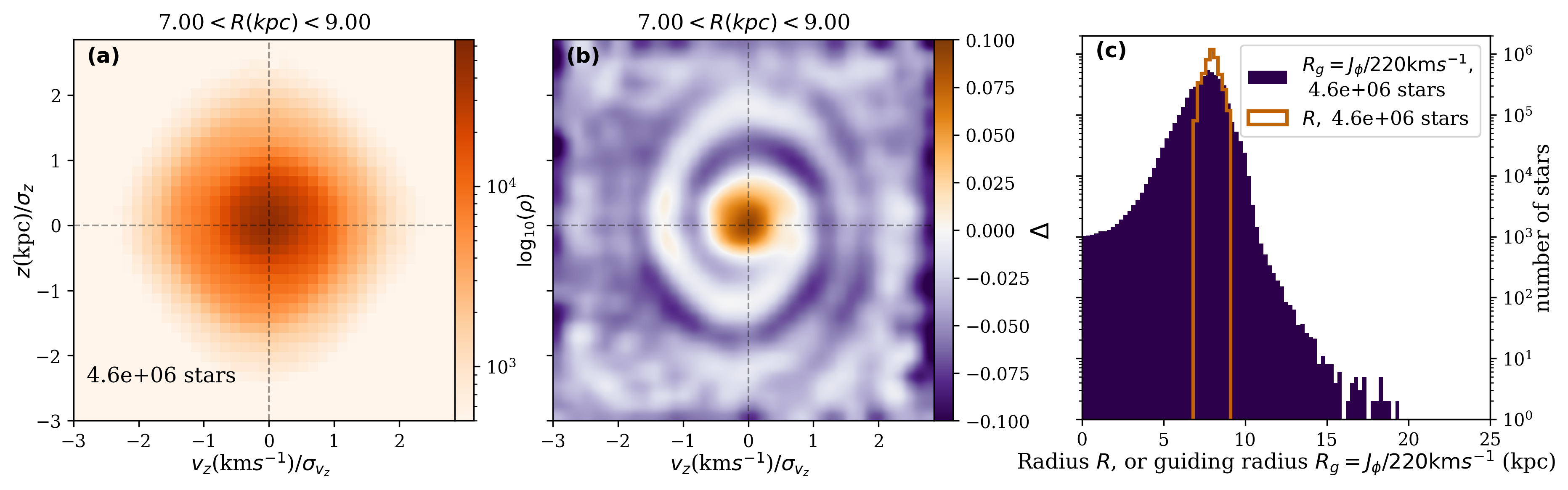}
    \caption{\textbf{(a)} The phase-space spiral as seen in the \z-\vz\ plane with data from \gaia\ eDR-3. We select $\sim4.6$ million stars within $7<$\rgal(kpc)$<9$ with parallax error $\sigma_\omega<20\%$. $v_z$ and $z$ have been rescaled by their respective dispersions, $(\sigma_{v_z},\sigma_z) = (25$  \kmpers,$0.37$  kpc).  The color represents the log number density. \textbf{(b)} Same as panel (a) except the color shows the filtered number density of \gaia\ eDR-3 stars, $\Delta\equiv (\rho-\overline{\rho})/\overline\rho$. Note that $\Delta$ is used specifically to highlight the spiral morphology, and it makes the spiral appear as a stream-like structure in phase-space. However, panel (a) is a more accurate representation of how the stars are actually distributed.   \textbf{(c)} A histogram of \rgal\ and $R_g\ (\approx J_\phi/220$ \kmpers) in our \gaia\ eDR-3 selection demonstrates that the \rgal-limited sample spans a much wider range in \rg.}
    \label{fig:gaia_all_data}
\end{figure}

\subsection{Dissecting One Phase-Spiral into Multiple} \label{subsec: dissect-snail}

\figref{fig:gaia_all_data}a shows the \zvz-plane for \gaia\ eDR-3 stars in the solar neighborhood, colored by the log number density. A phase-space spiral is visible in this local sample. 
\figref{fig:gaia_all_data}b is the same as \figref{fig:gaia_all_data}a, except the color bar represents the fractional overdensity relative to the mean number density ($\overline{\rho}$) at each pixel, $\Delta\equiv (\rho-\overline{\rho})/\overline{\rho}$ \citep[following][]{laporte2019footprints}.

We can dissect this sample further by exploiting the fact that stars which end up within the solar neighborhood today did not always travel together within the same enclosed volume. \cite{hunt2020} demonstrated how grouping stars around the disk by azimuthal action $J_\phi$ (equivalent to the $z$ component of the angular momentum $L_z$ in an axisymmetric potential) rather than radius \rgal\ more clearly separates them into sets that have shared histories. (For global disk samples, further grouping by the angle, $\theta_\phi$ conjugate to $J_\phi$, rather than physical angle $\phi$ can add further clarity to this separation.) The orbits of stars with similar $J_\phi$ can be characterized by epicyclic oscillations around the same guiding radius, $R_g$, and are  hence associated in space. Moreover, since they have the similar $R_g$, they also have similar azimuthal periods, and hence remain associated over time.  The guiding radius of a star is calculated by solving $R_g=J_\phi/v_c(R)$, where $v_c($\rgal) is the circular velocity as a function of \rgal. Throughout the paper, we assume a perfectly flat rotation curve with $v_c(R)\equiv v_0 = 220$  \kmpers, and estimate $R_g \approx$ \jphivo. This approximation of a constant $v_0=220$ \kmpers has been made for simplicity, and therefore the \rg\ values used in our work will not be exact. However, none of our results will be affected by this assumption. \figref{fig:gaia_all_data}c is a histogram of \rgal~(orange) and \rg\ (navy blue, filled) of the selected \gaia\ eDR-3 stars and clearly illustrates that although the local sample is limited by physical distance from the Sun, \rg\ allows us to probe a much wider radial range across the disk ($\sim$0--15 kpc, in this case).

\begin{figure}
\centering 
\subfloat[\textbf{Top panel:}  \zvz\ data of the local volume of stars  taken from \gaia\ eDR-3 ($7<R$(kpc)$<9$, same as \figref{fig:gaia_all_data}b). \textbf{Bottom row:} \gaia\ eDR-3 stars in the top panel now split into 5 \rg\ groups ranging $4<$ \rg (kpc)$< 12$. In all panels, we have rescaled \z\ and \vz\ by the respective dispersions $(\sigma_{v_z},\sigma_z)= (25$ \kmpers, $0.37 $kpc) to adjust the aspect ratio of the spirals. It becomes clear that a \rg\ categorization resolves distinct phase spirals (each panel in the bottom row shows a different spiral morphology) which otherwise get averaged out in the \rgal\ selection (top panel).]{%
  \includegraphics[width=\textwidth]{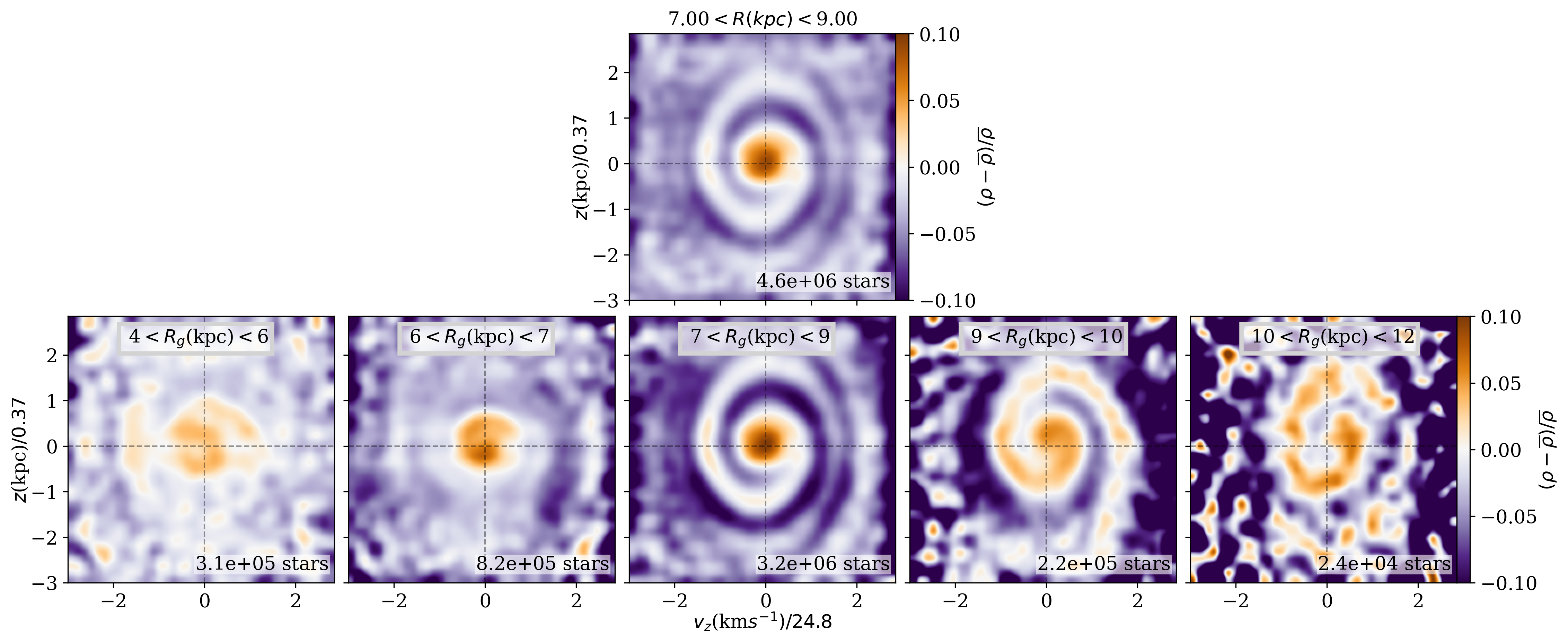}%
  \label{fig:gaia-jphi-covers-more}%
}\qquad
\subfloat[\textbf{Top panel:}  \zvz\ data of a local sample of stars ($6<R$(kpc)$<8$) in Model A of our test particle simulation. \textbf{Bottom row:} Simulated stars in the top panel now split into 5 \rg\ groups ranging 3 kpc$< \rg < 11$  kpc. In all panels, we have rescaled \z\ and \vz\ by the respective dispersions $(\sigma_{v_z},\sigma_z)= (27.6$ \kmpers, 0.36 kpc) to adjust the aspect ratio of the spirals. Once again, as is the case with the \gaia\ phase-spiral, $\rg$ categorization (bottom row) resolves distinct phase spirals which are averaged out in the test particle simulation \rgal\ selection (top panel). Colors outlining the \rg\ ranges specified in each panel (cyan, green, purple, magenta, red) and lower-case roman numerals (\textbf{(i)}--\textbf{(v)}) are identifiers used in figures throughout \S\ref{sec: results-I} to refer to the various \rg\ groups.]{%
  \includegraphics[width=\textwidth]{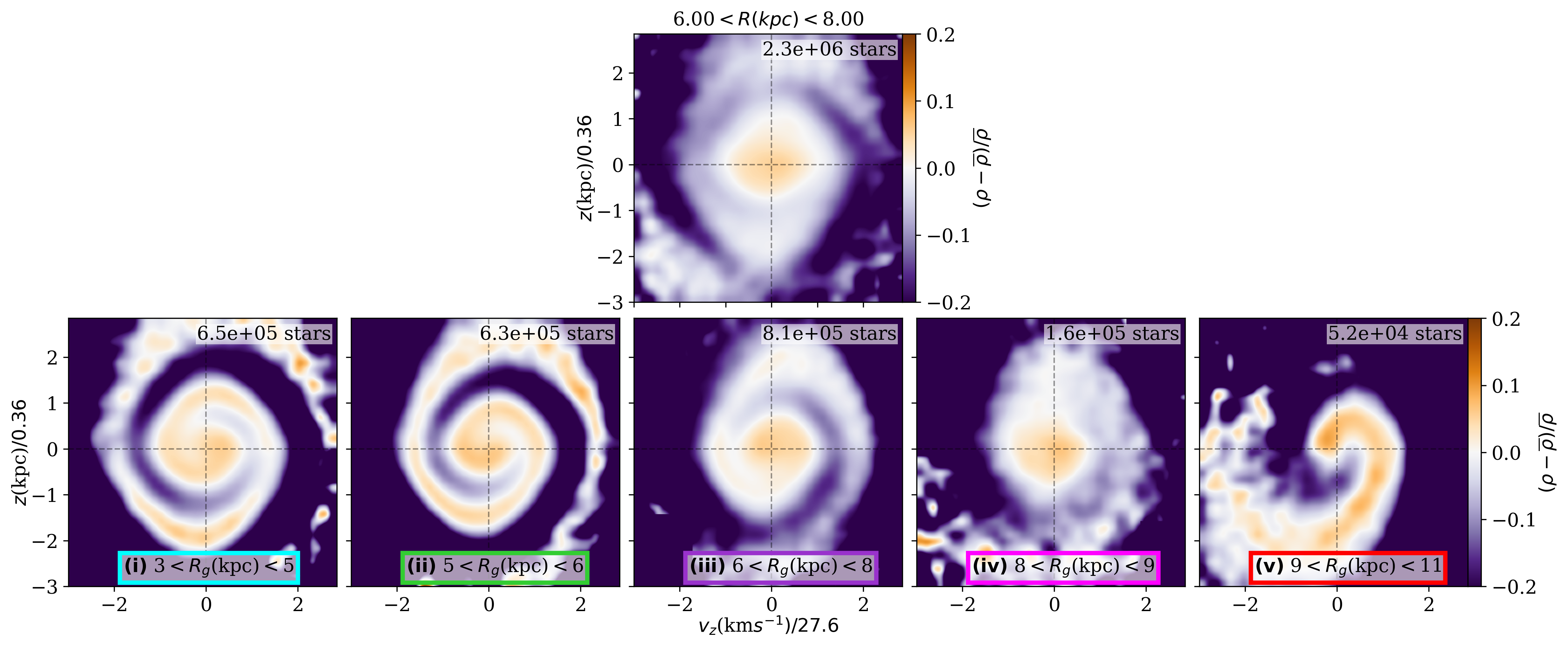}%
  \label{fig:sim-jphi-covers-more}%
}
\caption{}\label{fig:jphi-covers-more}
\end{figure}

In the top panel of \figref{fig:gaia-jphi-covers-more}, we repeat the same \zvz\ visualization of \gaia\ data as \figref{fig:gaia_all_data}b, and split this sample into five  \rg\ ranges in the bottom panels, between 4--6  kpc, 6--7  kpc, 7--9  kpc, 9--10  kpc, and 10--12  kpc. 
 
The vertical velocities and positions in each panel are scaled by their respective dispersions to adjust the aspect ratio of the phase spiral. We see from the bottom panels that \textit{distinct morphologies} of the spiral exist at different \rg~\textit{within the local sample} \citep[see also][]{li2020}. 

To perform an analogous split in test particle simulations, we select  a 30$^\circ$ azimuthal wedge in the disk between $6<$\rgal(kpc)$<8$ at a time chosen simply by virtue of the fact that it exhibits phase spirals with similar wrapping ($\lesssim2$ wraps), and significant variation across \rg, like in \figref{fig:gaia-jphi-covers-more}. This time happens to be $\sim$180 Myr after the second passage of the satellite galaxy through the disk, and we will refer to this as the ``sample" time. 
\figref{fig:sim-jphi-covers-more} shows rescaled \zvz\ for simulated stars within the specified local volume in the top panel, and the bottom panels show rescaled \zvz\  data for the \rgal-limited simulation sample split into five \rg\ bins, 
between \textbf{(i)} 3--5  kpc, \textbf{(ii)} 5--6  kpc, \textbf{(iii)} 6--8  kpc, \textbf{(iv)} 8--9  kpc, and \textbf{(v)} 9--11  kpc.
We use these five \rg\ bins throughout the remainder of \S\ref{sec: results-I}. 
As with the \gaia\ data, each \rg bin in the test particle simulation also reveals a different phase-spiral morphology, which otherwise gets obscured in the $6<$\rgal(kpc)$<8$ categorization.

    

\subsection{Tracing the Evolution of the  $R$--$\phi$ and \zvz\ Spirals by Rewinding the Test Particle Simulations}\label{subsec: tracing_back_spirals}

The fact that the morphology of the \zvz\ phase spiral within the local volume depends on  guiding radius raises the prospect of using this variation to study Galactic history. In order to explore the utility of the varied morphologies, we track  particles in the five \rg\ bins introduced in \figref{fig:sim-jphi-covers-more} backwards over time. We analyze their projections first in the \xy\ plane (forming \rphi\ spirals), and then in \zvz\ (forming phase-spirals).


\subsubsection{Physical Spirals: Global Phase-mixing in $R-\phi$} \label{subsubsec: x-y phase mixing}

\begin{figure}
    \centering
    \includegraphics[width = \textwidth]{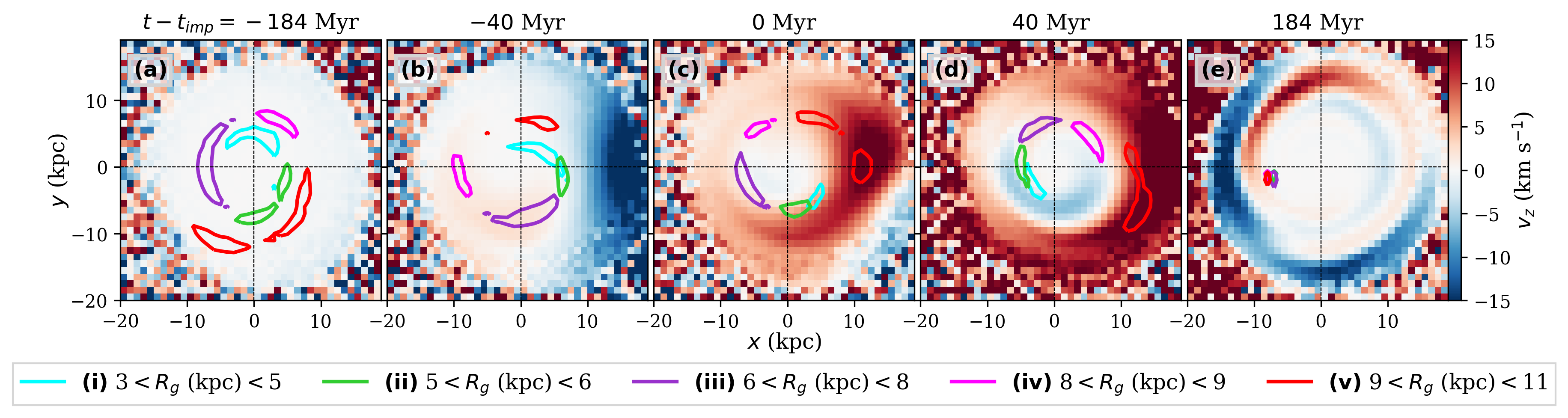}
    \caption{The 5 simulated \rg\ groups (shown in Figure \ref{fig:sim-jphi-covers-more}) are represented here by contours enclosing 50\% of the stars in each \rg\ bin, overplotted on the disk $x$--$y$ plane colored by $v_z$. Each panel shows the positions and velocities of simulated stars at a different time around the time of ``impact", $t=\timp$ (i.e., the time when the satellite crosses the disk midplane). \textbf{(a)} The \xy\ plane 184 Myr before \timp\ shows how different \rg\ groups are spread out across the disk. \textbf{(b)} 40 Myr before impact. \textbf{(c)} at the time of impact, $t=\timp$. \textbf{(d)} 40 Myr after the impact. \textbf{(e)} The ``sample" snapshot, 184Myr after the impact when all 5 \rg groups merge into a local volume at $\phi\approx-170^\circ$. This figure makes the point that two physical spirals can be identified extending across the disk---one that winds up as we go forward in time, highlighted by the \vz\ color in panels \ref{fig:5-jphi-groups-xy-plane-at-5-snaps}(c--e), and another that winds up as we go backward in time, traced by the various simulated \rg\ groups progressing from panel \ref{fig:5-jphi-groups-xy-plane-at-5-snaps}d back through \ref{fig:5-jphi-groups-xy-plane-at-5-snaps}a.
    The time intervals between the panels were chosen to be non-uniform (although symmetric about $t-\timp = 0$) so that the we can visualize the disk response long before/after the impact ($t-\timp = \pm 184$ Myr), as well as when the satellite is close to the disk ($t-\timp = \pm 40$ Myr). }
    \label{fig:5-jphi-groups-xy-plane-at-5-snaps}
\end{figure}

\figref{fig:5-jphi-groups-xy-plane-at-5-snaps} illustrates the evolution of two types of $R$--$\phi$ spirals, one that winds up as we go forward in time, and another that winds up as we go back in time. The figure shows snapshots starting at 184 Myr prior to the disk passage (panel \ref{fig:5-jphi-groups-xy-plane-at-5-snaps}a), through the time of the satellite impact, \timp~ (panel \ref{fig:5-jphi-groups-xy-plane-at-5-snaps}c), to the sample time 184 Myr after the disk passage (panel \ref{fig:5-jphi-groups-xy-plane-at-5-snaps}e). The first \rphi\ spiral is traced by the color across the face of the disk, which represents the mean \vz\ of all the particles. The influence of the satellite from under the disk pulling the particles downwards (blue color, $\vz<0$) and subsequently upward after crossing the midplane (red color, $\vz>0$) can be seen in the three middle panels (\ref{fig:5-jphi-groups-xy-plane-at-5-snaps}b--d). The global response eventually winds up (i.e. phase-mixes) into a clear spiral across the \xy\ plane by the ``sample" time. Note that this simple description of the satellite's impact followed by phase-mixing misses the additional effect present in reality and in the test particle simulation,  that the disk particles are also oscillating vertically. We will return to this in the next section.

The colored contours projected onto each panel trace the evolution of the second $R$--$\phi$ spiral. Each contour encloses 50\% of the stellar population in one of the five \rg\ groups shown in \figref{fig:jphi-covers-more}b. In panel \ref{fig:5-jphi-groups-xy-plane-at-5-snaps}a, the five groups (starting from the lowest \rg\ group in cyan, increasing through green, purple, magenta, and red), trace a tightly-wound physical spiral. The variation in azimuthal frequencies (see \omegaphi(\rg) in \figref{fig:freqs_and_time_periods}a) causes this spiral to unwind in each of the subsequent panels until all the groups coincide at the ``sample" time in the rightmost panel. It is striking to see in panel \ref{fig:5-jphi-groups-xy-plane-at-5-snaps}c that at the time of disk-crossing ($t=t_{\text{imp}})$, the different groups are spread widely across the disk in azimuth and radius. 


We conclude that, because the now-local stellar population was spread across the disk in the past, the \zvz\ spiral in each local \rg\ group contains distinct information about any past perturbation. We further point out that \textit{any} local volume in the disk will contain multiple phase-spirals, and the physical spiral in \vz across the \xy-plane is a signature of these multiple viewpoints averaged together.



\subsubsection{Phase-Spirals: Local Phase-mixing in \zvz} \label{subsubsec:z-vz phase mixing}

\begin{figure}
    \centering
    \includegraphics[width=\textwidth]{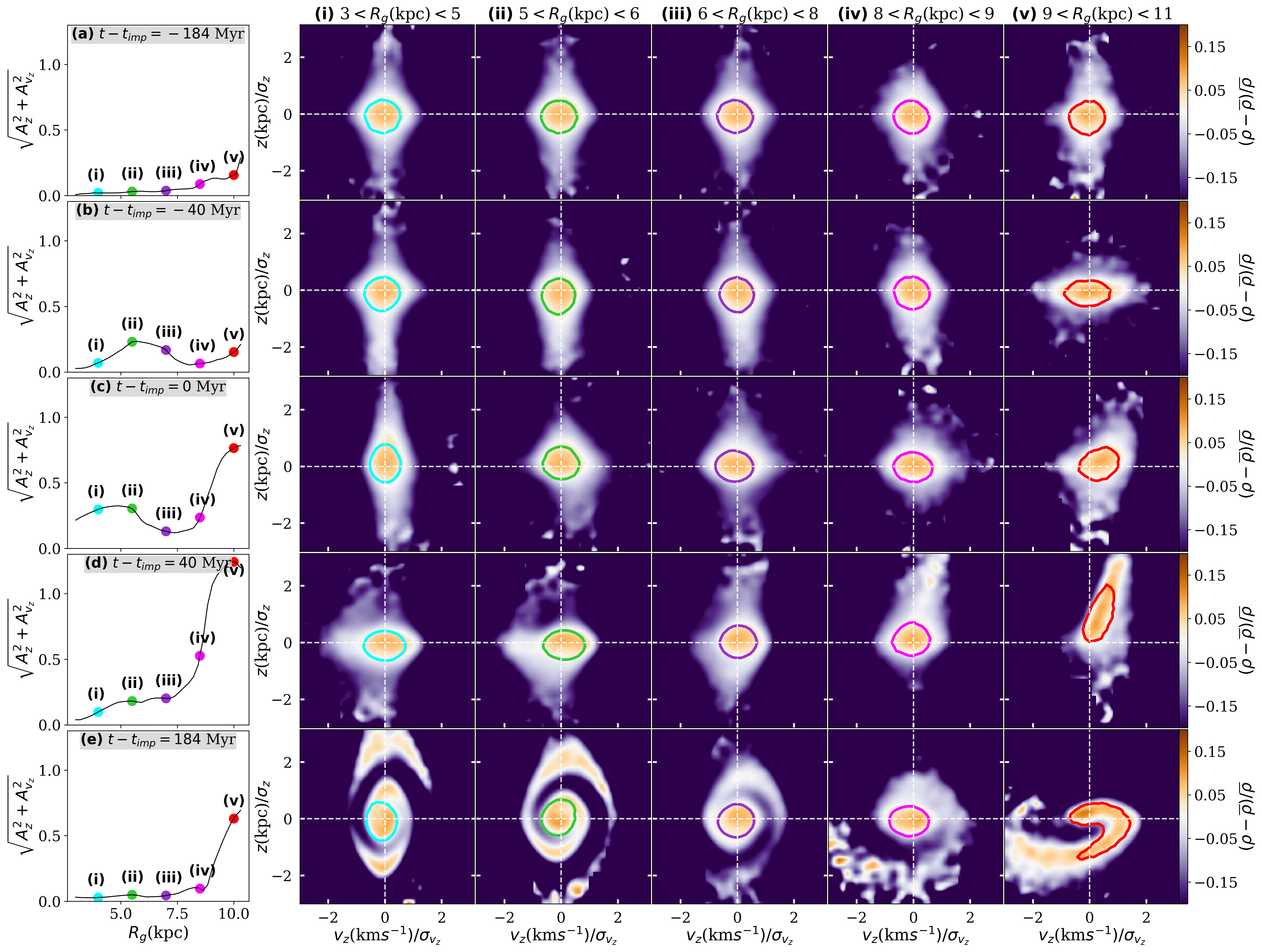}
     \caption{Each row \textbf{(a)}--\textbf{(e)} corresponds to a specific time relative to \timp (indicated in the leftmost panels), with the same time intervals as between the panels of \figref{fig:5-jphi-groups-xy-plane-at-5-snaps}. The leftmost panel in each row shows the \zvz\ asymmetry parameter, \azvz$\equiv \sqrt{A_z^2+A_{v_z}^2}$ (explained in the discussion around eq(\ref{eq: asym_param})) as a function of \rg. The five plots to the right (\textbf{(i)}--\textbf{(v)}) in each row show the five simulated \rg\ groups (introduced in \figref{fig:sim-jphi-covers-more}) in the \zvz plane. The colored contours in the \zvz\ panels enclose 50\% of the stellar population in each \rg\ group, and the \zvz\ asymmetry \azvz of each group is marked by a dot of the corresponding color and lower case roman numeral in the leftmost panel. \textbf{(a)} \azvz and the simulated \rg\ groups at $t-$\timp$=-184$ Myr. \azvz\ is minimal at this time. Most of the stars in all groups are symmetrically distributed in \zvz, as is evident from the 50\% contours being circles centered on $(v_z=0,z=0)$. This row corresponds to \figref{fig:5-jphi-groups-xy-plane-at-5-snaps}a. \textbf{(b)} At $t-\timp=-40$ Myr (corresponding to \figref{fig:5-jphi-groups-xy-plane-at-5-snaps}b), \azvz begins to grow. The 50\% contours are slightly distorted due to the disk's response to the approaching satellite.  \textbf{(c)} $t=\timp$: \azvz\ grows significantly for group \textbf{(v)} (red) as it is the one nearest to the location where the satellite crosses the disk (see \figref{fig:5-jphi-groups-xy-plane-at-5-snaps}c). \textbf{(d)} $t-$\timp$=40$ Myr (corresponding to \figref{fig:5-jphi-groups-xy-plane-at-5-snaps}d). \textbf{(e)} $t-$\timp$=184$ Myr, the ``sample" snapshot (corresponding to \figref{fig:5-jphi-groups-xy-plane-at-5-snaps}e).  In rows \textbf{(c)}--\textbf{(e)}, it is apparent that spirals develop at different rates for each \rg\ group. The asymmetry parameter \azvz oscillates over time, and is an important indicator of the amplitude of the response, which peaks even before any \zvz\ spirals appear. \azvz\ does not reflect how developed (or wound-up) a \zvz\ spiral is, but rather is a metric for asymmetric distribution of stars about $(v_z=0,z=0)$. Note that \z\ and \vz\ for  all \rg\ groups have been rescaled by the dispersions at $t-$\timp$=-184$ Myr. These are given by $\sigma_z=\{0.41,0.44,0.46,0.51,0.56\}$ kpc and $\sigma_{v_z}=\{44.5,37.4,34.3,29.8,23.4\}$  \kmpers for the five \rg\ groups respectively.}\label{fig:5-jphi-groups-zvz-plane-and-asym-param}
\end{figure}

Having tracked the \xy\ location of the five  \rg\ groups over time, we now explore the evolution of their morphologies in the \zvz\ plane.
As noted in \S\ref{sec:intro}, while the variation in \omegaz\ with \jphi\ (or \rg) causes a \textit{variety} of spiral morphologies to be apparent in the same local volume, it is the spread in \omegaz~ at a certain \jphi~ that leads to the phase spiral itself.
\figref{fig:freqs_and_time_periods}b illustrates both of these points with a 2D histogram of \omegaz\ as a function of \jphi\ for the simulated disk stars. The median \omegaz\ changes with \jphi, and at any given \jphi there is a range ($\Delta$\omegaz ) of values present. 
For instance, at $\sim 10$ kpc, the 1$\sigma$ dispersion about the median $\sigma_{\Omega z}\approx$ 2 epicyclic orbits per Gyr (marked by the dashed orange curves in \figref{fig:freqs_and_time_periods}b). Hence, over a few hundred Myrs we would expect particles with \rg$\sim$10 kpc that are displaced vertically by some perturbation to wind up, with the faster oscillating stars at the head of a \zvz\ spiral, and the ones with lower \omegaz~forming the tail.
These spirals will fade when particles have had time to fully phase-mix.
They will form more slowly in coordinates where stars have smaller spreads in frequencies.

In order to trace the onset and scale of responses in  \zvz\ in the simulated \rg\ groups, we adopt a simple \textit{asymmetry parameter} \citep[based on that used in][]{widrow2012, bennett_bovy2021}.  
The asymmetry ($A_X$) in a phase-space property $X$ of a group of stars is given by
\begin{equation} \label{eq: asym_param}
    A_X = \frac{N(X\geq0)-N(X\leq0)}{N(X\geq0)+N(X\leq0)}, ~~~~~~~~~~~A_X\in \{-1,1\}
\end{equation}
where $N(X\geq0)$ ($N(X\leq0)$) is the number of stars which have property $X\geq0$ ($X\leq0$).
For our purpose of detecting asymmetry specifically in the \zvz~plane, we introduce a combined kinematic asymmetry parameter $A_{z,v_z} \equiv \sqrt{A_z^2 + A_{v_z}^2} ~\in \{0, \sqrt{2}\}$.

Each row (\textbf{(a)}--\textbf{(e)}) in \figref{fig:5-jphi-groups-zvz-plane-and-asym-param} corresponds to a specific time relative to \timp\ (same time instances as the panels in \figref{fig:5-jphi-groups-xy-plane-at-5-snaps}). In each row, the leftmost panel has the time marked, and shows \azvz\ as a function of \rg; five panels  to the right (\textbf{(i)}--\textbf{(v)}) show the \zvz\ plane for each of the five simulated \rg\ groups at the respective time. Colored contours enclose 50\% of the stellar sample in each \rg\ group. Note that these contours do not necessarily enclose the \textit{same stars} as the contours in \figref{fig:5-jphi-groups-xy-plane-at-5-snaps}. Well before the disk crossing at $t-$\timp $= 184$ Myr (row \ref{fig:5-jphi-groups-zvz-plane-and-asym-param}a), simulated stars in all \rg\ groups are unperturbed and almost symmetrically distributed in the \zvz\ plane (i.e., the 50\% contours are circles centered on (\vz=0, \z=0) and \azvz(\rg)$\approx0$). By $t=\timp$ (\figref{fig:5-jphi-groups-zvz-plane-and-asym-param}c), the 50\% contours are distorted and \azvz\ has become large (especially for group \textbf{(v)} (red) as it is the one closest to the point of impact; see \figref{fig:5-jphi-groups-xy-plane-at-5-snaps}c). We see in rows \ref{fig:5-jphi-groups-zvz-plane-and-asym-param}(c--e) that the phase-spirals develop at different rates for the various \rg\ groups. \azvz oscillates over the course of the disk crossing. We emphasize here that \azvz\ is \textit{not} a measure of how developed (or wound) the phase spirals are, rather, it is a metric for how asymmetrical the distribution of stars is about (\vz=0,\z=0) .

\begin{figure}
    \centering
    \includegraphics[width = \textwidth]{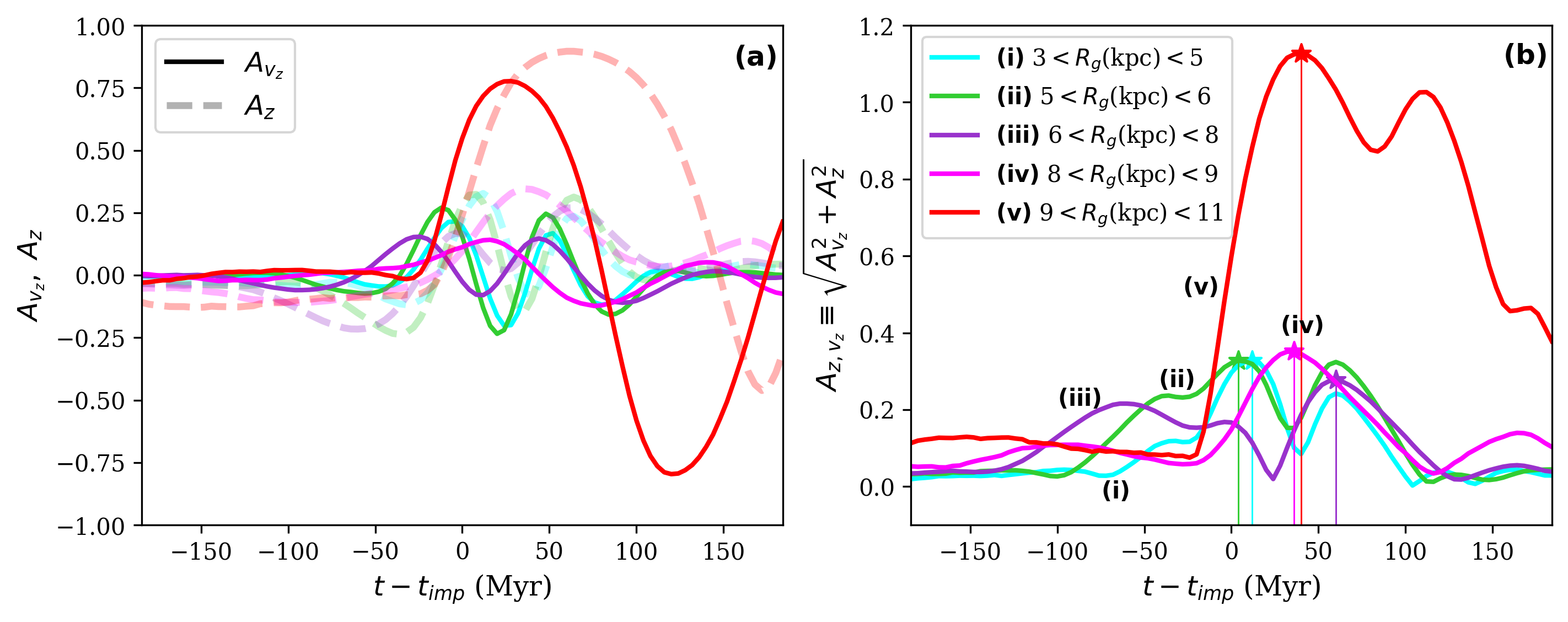}    \caption{Asymmetry parameters as a function of $t-$\timp\ for the five simulated \rg\ groups. For details about the different asymmetry parameters shown, see discussion around eq(\ref{eq: asym_param}). \textbf{(a)} $A_{v_z}(t-$\timp) (solid lines) and $A_z(t-$\timp) (dashed lines). Asymmetry oscillates with longer time period for groups with larger \rg. \textbf{(b)} \azvz$(t-$\timp) for the simulated \rg\ groups. The colored vertical lines topped with a `$\star$' mark the time of maximum \azvz\ for the \rg\ groups, and the maxima occur over a span of $\gtrsim 50$ Myr. These panels reiterate that stars in different regions of the disk experience the same perturbation (a satellite disk-crossing, in the case of our simulations) differently because the amplitude and oscillation frequency of asymmetry varies with \rg. The most important takeaway from this figure is that the effects of the disk-crossing begin to appear $\sim100$ Myr before \timp, and last well after \timp\ as well. Even though spirals don't appear until much later, the disk is differentially warped by the satellite's pull for $\gtrsim300$ Myr over the entirety of a single passage. }
    \label{fig:asymmetry_parameters}
\end{figure}

In \figref{fig:asymmetry_parameters}, we show $A_{v_z}$, $A_z$, and \azvz as a function of $t-$\timp~for each of the five simulated \rg\ groups. The vertical oscillations of the different groups are captured in \figref{fig:asymmetry_parameters}a, with the period of oscillation being longer for larger \rg\ (as seen in \figref{fig:freqs_and_time_periods}c). \figref{fig:asymmetry_parameters}b attempts to capture the growth of the response over time by plotting the amplitude of \zvz~ distortions overall. As expected, \azvz$\approx0$ for all \rg\ at early times. We assume \azvz$\sim \mathcal{O}(0.1)$ indicates the onset of the satellite's influence, which becomes apparent as early as $\sim100$ Myr before \timp. The vertical lines  with `$\star$' symbols mark the time of the maximum kinematic asymmetry for a particular \rg\ group, which corresponds to the time just before the spiral has wound up enough that the asymmetry starts to decline because of phase-mixing. This can occur up to $\sim50$ Myr after \timp, demonstrating the fact that responses can vary significantly with \rg. 

 A striking feature of \figref{fig:asymmetry_parameters} is the large asymmetry amplitude of the outermost \rg\ group ($\mathbf{(v)}\  9<\rg \mathrm{(kpc)}<11$, red curves) compared to the other groups. There are two contributing factors which explain this effect: (1) this \rg\ group is the one closest to the disk crossing region (red, $\vz>0$) at $t=$\timp\ (see \figref{fig:5-jphi-groups-xy-plane-at-5-snaps}c), and (2) the vertical oscillation period $T_z\approx 100$ Myr at $\rg\approx10$ kpc (see Figure \ref{fig:freqs_and_time_periods}; \omegaz$/2\pi(\rg=10\text{ kpc}) = 10$ Gyr$^{-1}$ or = 10 epicycles per Gyr), is approximately a third of the timescale of the disk crossing ($\sim300$ Myr--- the time over which the satellite causes significant asymmetry, estimated from \figref{fig:asymmetry_parameters}), thus leading to an enhanced, resonant response.  
 

Overall, we conclude that not only does each \rg\ group experience the interactions from a different location in the disk, but the interactions for each group also occur at different times and with different durations. Moreover, the interaction is far from impulsive, but rather comparable to the orbital times.

\subsection{Toy Model of the Influence of a Satellite During a Disk Crossing} \label{subsec: toy_model}

In this section, we use a toy model of the influence of a satellite crossing the disk in order to place the results of prior sections in context --- how the experience of the same satellite perturbation can vary across \rg\ groups.
For the purpose of isolating the scale of the influence in different regions of the disk, we assess the impact on toy model stars moving in the midplane (\z$_\star$\ fixed to 0) on perfectly circular orbits ($\hat v_\star$ fixed to $\hat\phi$) throughout the encounter and ignore their vertical oscillations or any vertical displacement due to the satellite's pull.

\begin{figure}[h]
    \centering
    \subfloat{%
    \includegraphics[width=0.49\linewidth]{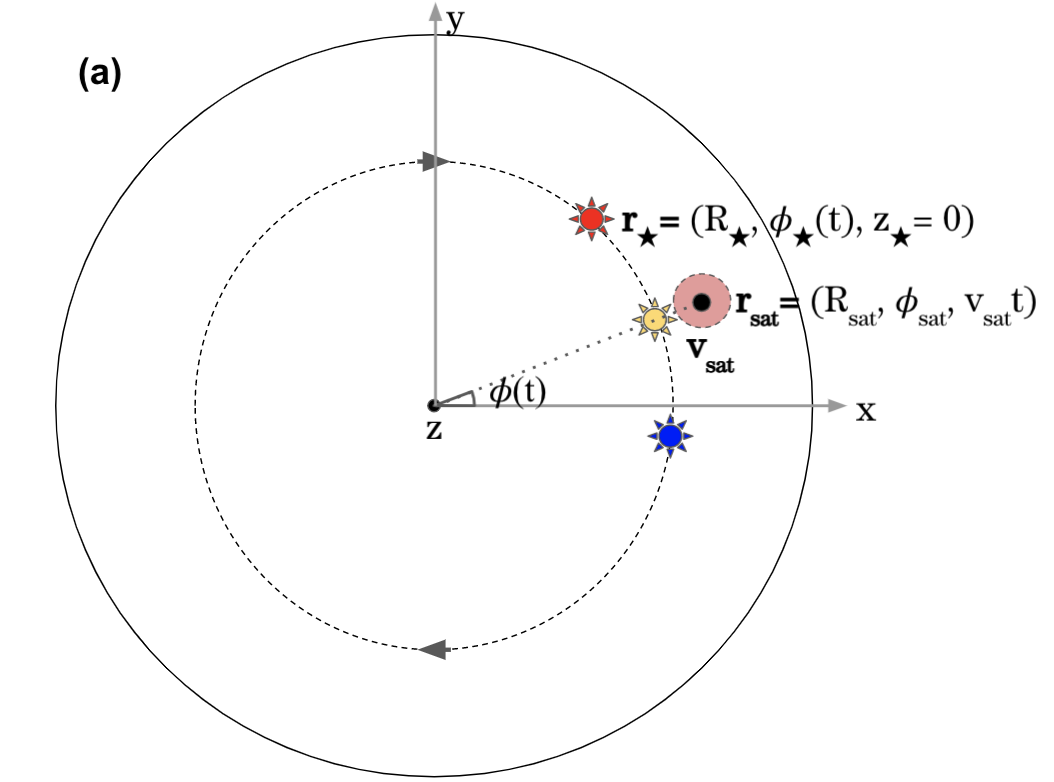}
              }
\subfloat{
        \includegraphics[width = 0.49\linewidth]{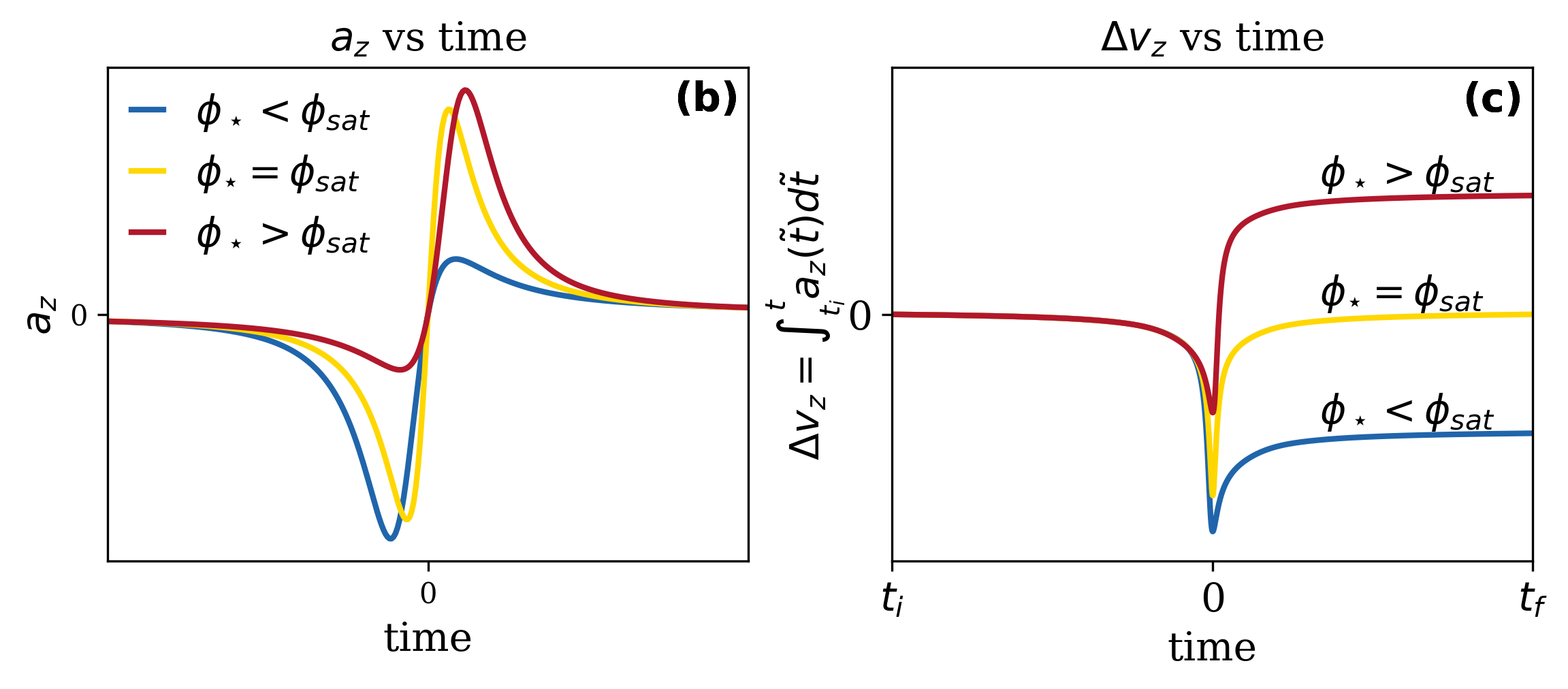}
}
    \caption{\textbf{(a)} The $x$-$y$ projection of a toy disk with a satellite (pink circle at $\vec{r}_{\text{sat} }= (R_{\text{sat}},\phi_{\text{sat}},v_{z,\text{sat}}t)$) passing through on a strictly vertical path with $v_{z,\text{sat}}>0$ (out of the page). The disk rotates clock-wise, with all stars remaining strictly in the midplane and at constant radius represented by $\vec{r}_\star = (R_\star,\phi_\star(t),z_\star=0)$. \textbf{(b)} Vertical acceleration caused \textit{solely} by the satellite, $a_z(t-\timp)$, where \timp\ is the time the satellite crosses the midplane, for the 3 toy model stars marked in panel (a). \textbf{(c)} Cumulative \az$(t-\timp$) for the 3 sample toy model stars. Panels (b,c) together show that over an extended time $t_i< t< t_f$ such that the initial time $t_i\ll $\timp\ and final time $t_f\gg\timp$ , the red star ($\phi_\star(\timp) > \phi_{sat}$) gets a net positive $\Delta v_z(t_f)$, the blue star ($\phi_\star(\timp) < \phi_{sat}$) gets a net negative $\Delta v_z(t_f)$, and the yellow star ($\phi_\star(\timp) = \phi_{sat}$) experiences a net zero change in $v_z$ at $t=t_f$. See discussion in \S\ref{subsec: toy_model} for further explanation. (\textit{Note:} The time axes in panels (b) and (c) are different; we have zoomed into a shorter time interval in panel (b) compared to (c) to show the $a_z(t)$ curves in better detail.)}
    \label{fig:impact_toy_model_schematic}
\end{figure}

\figref{fig:impact_toy_model_schematic}a sketches the toy model: a galactic disk (shown face-on) comprises stars on clockwise circular orbits in the midplane, and a satellite on a vertical trajectory passing through it with \vz$>0$ (out of the page). 
We can get some intuition for our results by first considering the three toy model stars shown in \figref{fig:impact_toy_model_schematic}a, one with $\phistar(\timp) = \phisat$ at the time of disk crossing (yellow star), a second with $\phistar(\timp) < \phisat$ (blue star), and the third with $\phistar(\timp) > \phisat$ (red star). \figref{fig:impact_toy_model_schematic}b demonstrates that the yellow star experiences exactly equal and opposite force from the satellite before and after \timp, and thus experiences net $\Delta v_z =0$. The blue star is closer to the satellite when it is being pulled down (at $t<\timp$), but farther when being pulled up (at $t>\timp)$ and thus has net $\Delvz<0$. Finally, the red star is farther from the satellite when it is being pulled down (at $t<t_{imp}$), and closer when being pulled up (at $t>t_{imp})$ and thus has net $\Delta$\vz$>0$. \figref{fig:impact_toy_model_schematic}c simply reiterates the asymmetric response by showing the cumulative \az$(t-\timp)$ which is 0 at $t-\timp = t_i\ (t_i\ll \timp)$ for the three sample toy model stars, and by $t-\timp = t_f\ (t_f\gg \timp)$, cumulative \az\ is positive for the red star ($\delphi>0$), $0$ for the yellow star ($\delphi=0$), and negative for the blue star ($\delphi<0$).

We quantify the scales of the satellite influence across the disk by integrating the  \z-acceleration it exerts on some sample particles over the course of the encounter. Each test particle has a position vector in cylindrical coordinates, $\vec{r}_\star = (R_\star, \phi_\star(t), z_\star = 0)$, such that its galactocentric radius is constant and it remains in the midplane. The azimuth of a particle is given by $\phi_\star = \phi(\timp) + v_\star t$. The satellite (pink circle) of mass $M_{sat}$ is on a vertical trajectory with constant galactocentric radius, azimuth, and upward vertical velocity, described by $\vec{r}_{\text{sat} }= (R_{\text{sat}},\phi_{\text{sat}},z_{\text{sat}} = v_{z,\text{sat}}t)$). 
The vertical acceleration $a_z(t)$ of a particle due to the gravitational pull of the satellite is given by,
\begin{equation} \label{eq: az}
    a_z(t) = G M_{sat} \frac{z_{sat}(t) - z_\star}{|\vec{r}_\star(t)-\vec{r}_{sat}(t)|^3} =  G M_{sat}\frac{v_{z,sat}t}{\left[R_{sat}^2+R_*^2 - 2R_{sat}R_*\cos{(\phi_*(t)-\phi_{sat})} + (v_{z,sat}t)^2\right]^{3/2}}
\end{equation}
Note that we are intentionally \textit{only} considering the acceleration due to the satellite (and not including the restoring force of the disk) because we want to isolate the response generated by the satellite. Thus in this paper, $a_z$ refers to the vertical acceleration defined in \eqref{eq: az}), and we neglect the evolution of orbits by forcing the toy model stars to remain on circular orbits.

\begin{figure}[ht]
    \centering
    \includegraphics[width=\textwidth]{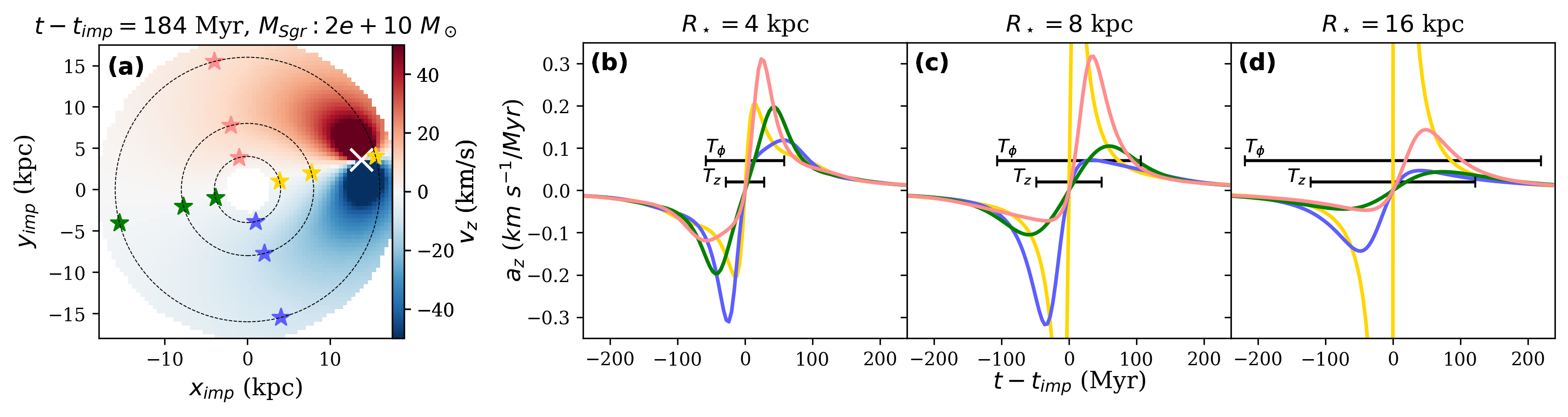}
    \caption{\textbf{(a)} A toy disk rotating clockwise in the $x$-$y$ plane colored by vertical velocity $\Delta v_z$ over $-184 <t-\timp$ (Myr)$<184$. The positions are frozen at $x_{imp},y_{imp}$ at $t=t_{imp}$. The white `$\times$' marks the point where the satellite crosses the midplane with \vz$_{\text{,sat}}>0$, coming out of the page. This panel demonstrates the point that toy model stars with $\phi_{\star}(\timp) > \phi_{sat}$ get a net positive kick in $v_z$ (red region), the toy model stars with $\phi_{\star}(\timp) < \phi_{sat}$ get a net negative kick (blue region), whereas toy model stars with $\phi_{\star}(\timp) = \phi_{sat}$ or $\phi_{\star}(\timp) = \phi_{sat}-\pi$ get a net $0$ change in $v_z$. For stars in this toy model, $\Delta v_z$ is calculated by only accounting for \az\ caused by the satellite passing through the disk (see Eq.(\ref{eq: az})). That is, we neglect epicyclic oscillations and self gravity. Three sample radii are chosen (black dotted circles at 4, 8, and 16  kpc) with sample stars (marked with `$\star$' symbols) distributed in azimuth to analyze $a_z(t)$. \textbf{(b)--(d)} For each sample radius (\textbf{(b)} 4 kpc, \textbf{(c)} 8 kpc, and \textbf{(d)} 16 kpc), we show $a_z(t-t_{imp})$ for the sample stars selected at that radius. The color of each curve in these panels corresponds to a colored `$\star$' marked in panel \ref{fig:az-lag-lead-3stars}a. The red curves have a net positive $\Delta$\vz, yellow and green curves have net zero, and blue curves have net negative $\Delta$\vz. The black horizontal bars indicate time periods of oscillations at the sample radius (taken from Figure \ref{fig:freqs_and_time_periods}a): the longer bar is $T_\phi$ (period of $\phi$-rotation), the shorter one is $T_z$ (period of vertical epicyclic oscillation). The vertical acceleration of the toy disk stars last over a period comparable to orbital time scales. \textit{Note:} The stars in this toy model are \textit{not} oscillating, the horizontal bars are simply shown to indicate what the oscillation time periods are in a MW-like disk at particular \rgal\  values, and how they compare to the time scale of the disk crossing.}
    \label{fig:az-lag-lead-3stars}
\end{figure}

We calculate $a_z(t)$ for all particles in a toy disk where we set $M_{sat} = 2\times10^{10}\ M_\odot$, $z_{sat}(t) = v_{z,sat}t$, $v_{z,sat} = 339$ km$s^{-1}$, $\phi_\star(t) = \phi_\star(\timp)+v_\star t$, and $v_\star(R_\star)$ is obtained from the rotation curve of \mwpot\ in \galpy. The satellite's mass, vertical velocity, and disk crossing coordinates are chosen to closely match the satellite in Model A of the test particle simulation at $t=$\timp. Over a certain time period $t_i\leq t\leq t_f$, we can find the total change in a toy model star's vertical velocity,
\begin{equation} \label{eq: delta_vz}
    \Delta v_z = \int_{t_i}^{t_f} a_z \, \textrm{d}t .
\end{equation}
\figref{fig:az-lag-lead-3stars}a shows a toy disk colored by $\Delta v_z$ integrated over $-0.5<t-\timp\text{ (Gyr)}<0.5$. The \xy\ coordinates of stars in this panel are frozen at \timp\ to show net $\Delta v_z$ at $t=t_f$ as a function of position at $t=\timp$ with respect to the satellite at disk-crossing. Toy model stars with $\delphi<0$ experience net \Delvz $<0$ over the entire course of the disk crossing and lie in the blue region, whereas stars with $\delphi>0$ have the opposite response and lie in the red region. Only at $\delphi=(0,\pi)$ is \Delvz$=0$. 

Three sample radii of \rgal$=(4, 8, 16)$  kpc with four sample toy model stars at each radius are chosen (shown with `$\star$' symbols on dashed circles in \figref{fig:az-lag-lead-3stars}a) to examine $a_z(t)$ for different radii and azimuths. Each of Figures \ref{fig:az-lag-lead-3stars}b-d shows $a_z(t-t_{imp})$ for each of the sample radii, and the colored curves corresponds to a `$\star$' in \figref{fig:az-lag-lead-3stars}a of the same color at that radius. The horizontal bars mark $T_\phi$ and $T_z$ for the corresponding \rgal. 
Note that epicyclic oscillations are not included in the toy model; $T_z$ is simply taken from the simulated data in \figref{fig:freqs_and_time_periods}a to compare typical oscillation timescales with the disk crossing timescale. 

The toy model captures the qualitative diversity in the disk response to a satellite perturbation. It demonstrates the \rgal-\ and \phistar-dependence of $a_z(t)$. It underscores the fact that the satellite's integrated influence on vertical motions: (i) can have opposite signs in different disk regions;  (ii) reaches its peak at different times for different azimuths at the same radius; (ii) extends over timescales comparable to both the orbital and vertical oscillation times. Of course, these conclusions are a function of the particular passage we choose to examine --- these are the characteristics we expect for satellites with disk crossings at radii that are comparable to the disk size itself. Another caveat of this simple model is that it neglects $\Delta v_R$ and $\Delta v_\phi$ of stars, which will qualitatively alter the signatures of vertical disturbance we see in the toy disk.


\subsection{Combined Implications of Our Results}\label{subsec: results_I_conc}

We have explored the origin of the multiple \zvz\ spiral morphologies apparent when local samples in the \gaia\ data and test particle simulations are divided into \rg\ bins. In Section \ref{subsubsec: x-y phase mixing} we showed how  the locations of these distinct samples, traced back in time,  outline an \rphi\ spiral across the disk (see distribution of colored contours in \figref{fig:5-jphi-groups-xy-plane-at-5-snaps}a). Furthermore, the various \zvz\ spiral morphologies in \textit{any} local volume are \textit{local} signatures of a \textit{global} \vz\ spiral spanning the extent of the disk (see \vz\ structure in \figref{fig:5-jphi-groups-xy-plane-at-5-snaps}e).  

In Section \ref{subsec: toy_model} we used a toy model to examine the case of a $2\times 10^{10} M_\odot$ satellite crossing the Galactic disk plane and demonstrated how the overall influence is neither local nor impulsive. Stars at different \rg\ which form phase spirals ``today", started responding to a perturbation tens of Myrs before \timp\ and well before coherent phase spiral structures developed. The combination of these results imply that the multiple local phase-spirals, even if associated with the same event, cannot be simply ``rewound" to a single time and location to learn about the impact. Rather, each \rg\ group represents distinct viewpoints  of the same event which are widely spread out in space and time.

We reiterate that although we present the example of a disk-and-satellite interaction as the cause of these phase-spirals, their subsequent phase-mixing in \zvz~ and \rphi\ depends only on the disk properties and not the nature of the interaction.  Distinct morphologies in local \zvz\ samples would occur with other types of perturbations as well, and contain multiple viewpoints on the cause, whether a buckling bar, rippling disk or an impacting satellite. Hence local spirals could be powerful diagnostic tools of global disk disturbances.

\section{Results II: Application to the Ongoing MW-Sgr Interaction}\label{sec: results-III_ongoing_interaction}

The results from previous sections show that a \sag-like satellite passing through the midplane at \rgal $\approx15$ kpc can influence the disk at times $\sim \pm 150$Myr around \timp. Currently, \sag~ is approaching the MW from $z\approx-6$ kpc with \vz $\approx200$ \kmpers and is expected to hit the outer disk at $R\sim 18$ kpc in $\sim$30Myr. This suggests that, even though \sag~has not yet crossed the midplane, there could be signatures of this encounter already developing in the disk.

\figref{fig:sgr_galpy_orbit} shows the recent and near future path of  \sag's by tracing its orbit within \galpy's \mwpot~forwards and backwards from its present-day Galactocentric phase-space coordinates from \cite{Vasiliev2020} noted in \S\ref{subsec: tp sim description}.
The geometry of \sag's orbit causes it to travel quite close under the disk plane as it approaches its present location rather than simply passing vertically. To explore possible signatures of the current interaction, we now analyze global deviations from zero in \vz\ as well as local phase-space signatures which might be detectable in future surveys.  
On these short timescales ($\sim$150 Myr), we expect disk self-gravity to be least important and our toy and test particle models to capture much of the  response.

\subsection{Estimates of Scale from the Toy Model}\label{subsec:toy_model_current_interaction}
\begin{figure}
    \centering
    \includegraphics[width=0.6\textwidth]{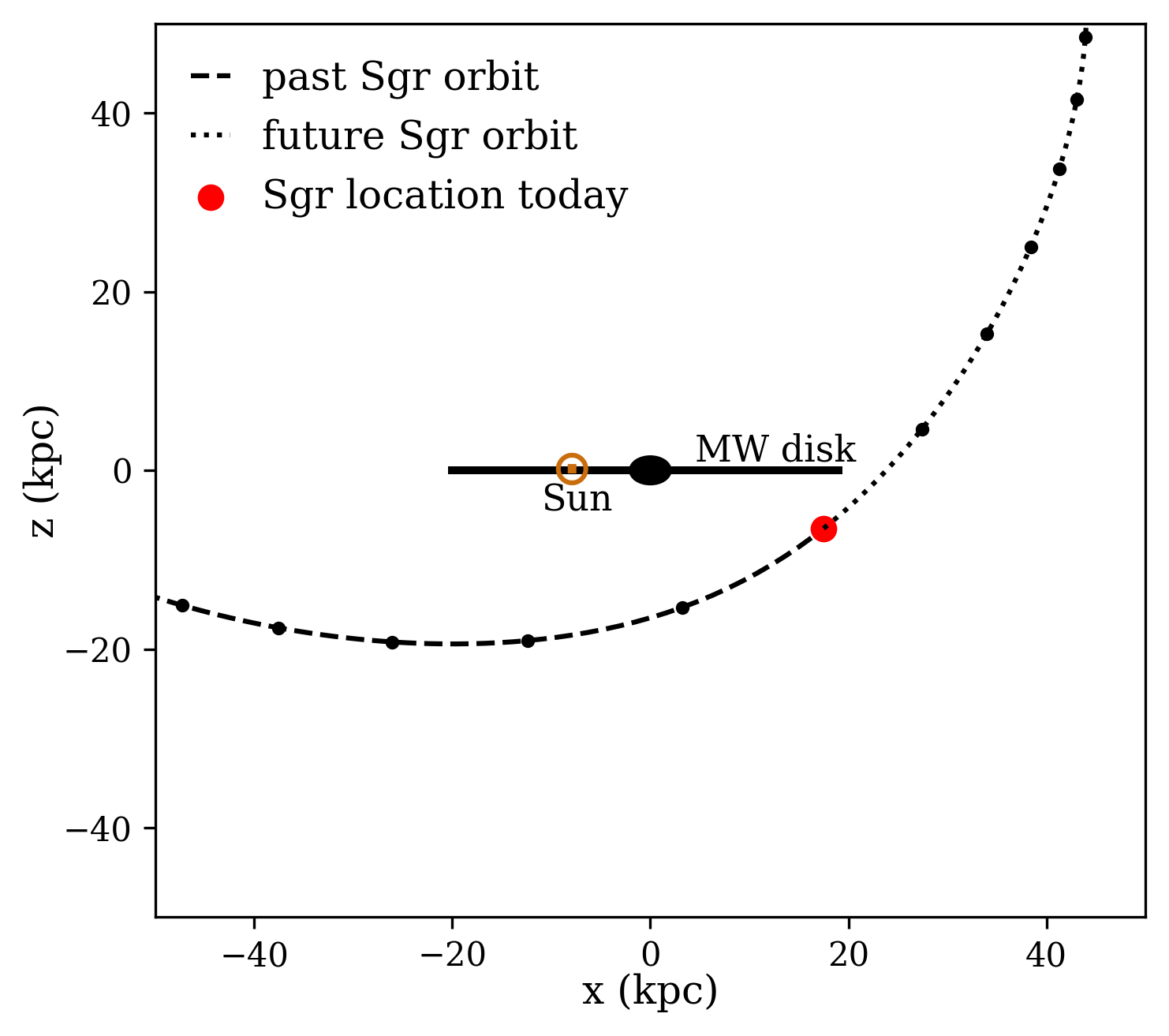}
    \caption{The $x$--$z$ projection of \sag's orbit with present day phase space information from \cite{Vasiliev2020}. The orbit has been integrated in \mwpot\ from \galpy. The current position of Sgr is marked by the large red dot, the dashed curve is the the past orbit, and the dotted curve is the future orbit. The black dots are 50 Myr apart. The black horizontal line represents the MW disk, with the Sun marked by the orange `$\odot$'. The asymmetry of Sgr's past and future orbit indicates that soon after crossing the midplane, its distance from the disk will increase significantly compared to the past $\sim150$ Myr while it was sweeping under the disk. Its proximity to the disk in the recent past leads us to explore whether signatures of vertical disturbance have already developed.}
    \label{fig:sgr_galpy_orbit}
\end{figure}

We apply our toy model to explore the nature of \sag's  past, present, and future influence on the MW throughout the current disk passage. The toy disk once again comprises particles on strictly circular orbits  rotating clockwise with constant circular velocity \vstar($R_\star$) and $z_\star =0$, but this time for a satellite travelling along the orbit whose $x--z$ projection is shown in  \figref{fig:sgr_galpy_orbit}. We set \Msg\ to a constant $3\tenexp9\ \msun$ \citep[the initial \sag\ mass in][]{Vasiliev2020}.

\begin{figure}
\includegraphics[width = 1\textwidth]{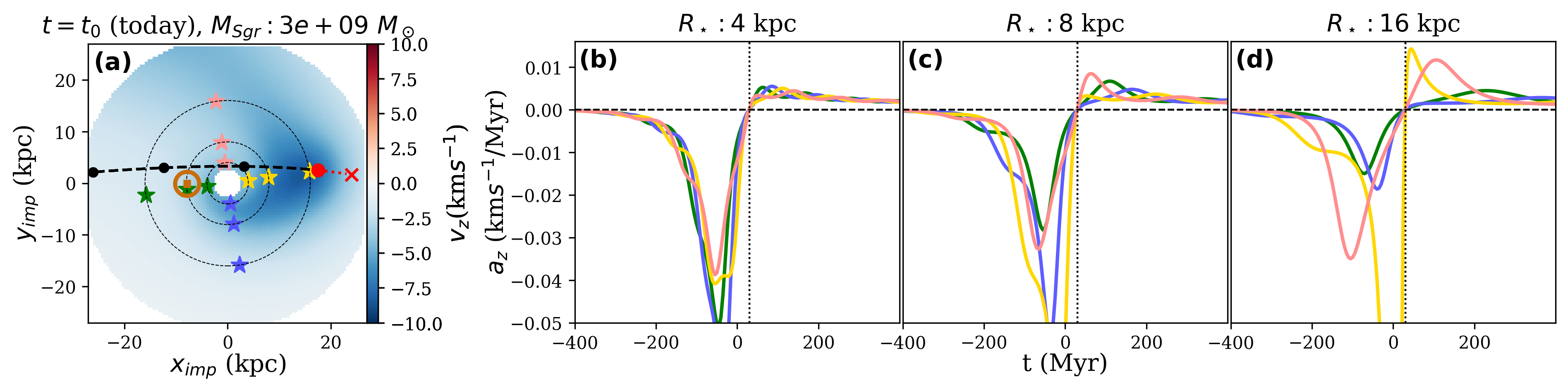}
\caption{This figure's panels and color schemes are the same as Figure \ref{fig:az-lag-lead-3stars}; $M_{Sgr} = 3\times10^9 M_{\odot}$. \textbf{(a)} The $x$-$y$ plane colored by $\Delta v_z$ integrated over the past 500 Myr up until present day. Since Sgr has been below the disk for the past $\sim200$Myr and is at $z\approx -6$ kpc at present, there is a net negative $\Delta v_z$ across the entire disk. The red dot shows the current $x$--$y$ position of Sgr, and the black dashed line tracks its past trajectory, with the black dots being 50Myr apart in time. The red dotted line tracks Sgr's future trajectory and the red cross marks the position where Sgr will cross $z = 0$ ($\sim$30 Myr in the future). The orange `$\odot$' at $(-8,0)$  kpc indicates the current position of the Sun. \textbf{(b)--(d) }$a_z(t)$ at different radii and azimuths, with $t=0$ at present day. The dotted vertical line marks the time in the future when Sgr will cross $z=0$. The extreme asymmetry of $a_z$ around $t=0$ in these three panels makes the point that a major fraction of the $\Delta v_z$ due to the imminent passage of Sgr \textit{has already been induced} by present day. Once again, this toy model does not account for epicyclic oscillations and self-gravity, therefore should not be interpreted as quantitatively accurate.} \label{fig: az_toy_model_Sgr_current_passage}
\end{figure}

\figref{fig: az_toy_model_Sgr_current_passage}a shows the disk \xy\ plane at present day $(t=t_0$, ``today"), colored by $\Delta v_z$ (see eq(\ref{eq: delta_vz})) integrated over $-0.5\leq t-t_0\text{\ (Gyr)}\leq 0$. The black dashed curve shows the \xy\ projection of \sag's past orbit, with the black dots spaced at intervals of 50 Myr. The large red dot marks \sag's current \xy\ position, the red dotted curve is its future orbit and the red cross is where it will cross $z=0$. The orange `$\odot$' at ($-8,0$) marks the Sun.  
The plot shows that particles across the disk receive a net velocity kick of up to several  \kmpers\ from this portion of the passage alone. Again, remember that the toy model neglects epicyclic oscillations and thus is not a prediction for the mean \vz, but rather for the scale of \sag's influence in different regions. Moreover, the mass of \sag\ that  we have used (\Msg$\approx3\tenexp{9}$\msun) is the initial and largest mass in \citet{Vasiliev2020} (see their Figure 9). The Sgr remnant loses significant mass by present day, which would mean that the strength of the \vz\ signal in reality will be much weaker than the \vz\ amplitude in our toy model.

Figures \ref{fig: az_toy_model_Sgr_current_passage}b--d each show \az$(t)$ for radii \rgal=4, 8, and 16  kpc respectively and particles at four different azimuths (`$\star$' symbols in \ref{fig: az_toy_model_Sgr_current_passage}a).  The asymmetry of \az$(t)$ around $t_0$ emphasizes the fact that most of the vertical perturbation within $\rgal\lesssim10$  kpc due to \sag's ongoing interaction with the Milky Way \textit{has already happened} over the past $\lesssim200$Myr, as the satellite sweeps close under the disk. More specifically, the ratio between $|\Delta v_z|$ induced over the past 200 Myr and $|\Delta v_z|$ induced over $t_0 \pm 200$ Myr , averaged over the 4 sample toy model stars at each radius is 0.82, 0.79, and 0.69 for $R_\star = 4,\ 8,\ 16$ kpc respectively.

\subsection{Morphological Predictions from Test Particle Simulations}\label{subsec: current_interaction_sim_prediction}

Our qualitative conclusions from the toy model in \S\ref{subsec:toy_model_current_interaction} motivate searching for a signature of the vertical response to the ongoing MW-\sag~interaction. We use test particle simulations of a disk galaxy perturbed by a Sgr-like satellite on the orbit prescribed by \mwpot~with present day Sgr phase-space coordinates from \citet{Vasiliev2020}. We analyze the simulations for two different cases--- \hyperref[tab:I]{Model B}: a MW-like disk which has only evolved under Sgr's influence over the past 150 Myr with \Msg$=3\tenexp9\msun$, and \hyperref[tab:I]{Model C}: a MW-like disk that has experienced multiple crossings of \sag\  over the past 4 Gyr, with \Msg$=10^{10}\msun$. This helps us extricate the vertical response over the past 150 Myr from the remnant effects of past disk crossings. It also allows us to see how the amplitude of the response scales with \Msg.

\subsubsection{Physical Spirals}
    \begin{figure}
    \centering
    \includegraphics[width=\textwidth]{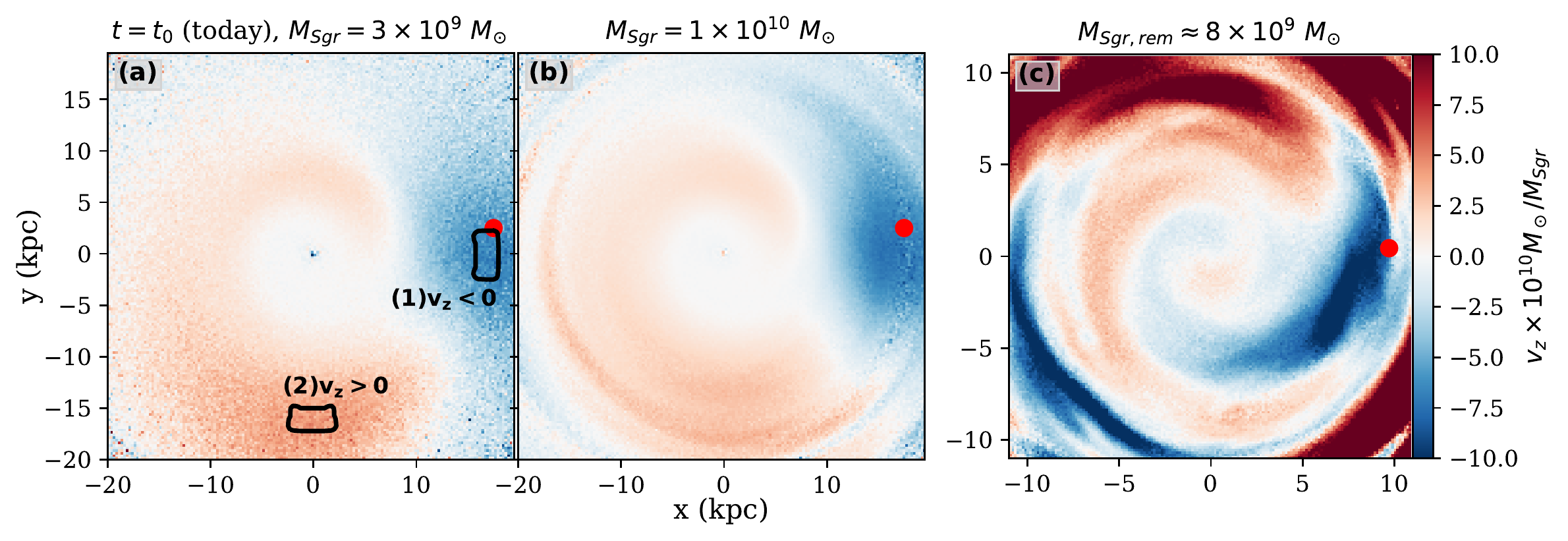}
    \caption{\textbf{(a)} Model \hyperref[tab:I]{B}: a test particle disk \xy\ plane at present day ($t=t_0$) colored by vertical velocity scaled by \Msg\,  $v_z \times 10^{10}M_\odot/M_{Sgr}$ (this rescaling helps enhance the \vz\ signal from a satellite with \Msg$<10^{10}\msun$). The test particle disk has evolved under the influence of \sag\ \textit{only} over the past 150 Myr with $M_{\text{Sgr}} = 3\times 10^{9}\ M_\odot$. The outlined patches correspond to the panels of \figref{fig: zvz_present_day_model_b}. \textbf{(b)} the same plot as (a), but for Model \hyperref[tab:I]{C}.  The test particle disk has evolved over the past 4 Gyr, and $M_{Sgr} = 10^{10}\ M_\odot$. The \vz\ pattern across the disk is largely the same between panels (a) and (b), and the \vz\ amplitudes are also similar once scaled by \Msg. The fact that the isolated disk in (a) shows the same \vz\ pattern as (b) implies that the current MW-Sgr interaction is largely what creates this \vz\ signature. \textbf{(c)} ``Present day" snapshot from a  self-consistent simulation (described in \S\ref{subsec: bonsai sim description}) for purely illustrative purposes to show that a similar pattern in \vz\ emerges even when self-gravity of the disk is accounted for. Note that the \xy\ limits of this plot are different from panels (a) and (b), which means that the extent of the \vz\ pattern is not the same. However, this can be explained by the various differences between the \sag\ orbit and disk potential between the test particle and self-consistent simulations. The main takeaway from this figure is that in \textit{all three panels,} a blue region with $v_z<0$ emerges around $\phi\sim0^\circ$, and a red region with $v_z>0$ is present around $\phi\sim-90^\circ$. This leads to a robust prediction that these signals in \vz\ are present in the MW disk today and might be detectable in future surveys. }
    \label{fig:xy_current_interaction_present_day_different_models}
\end{figure}

\begin{figure}
    \centering
    \includegraphics[width=0.7\textwidth]{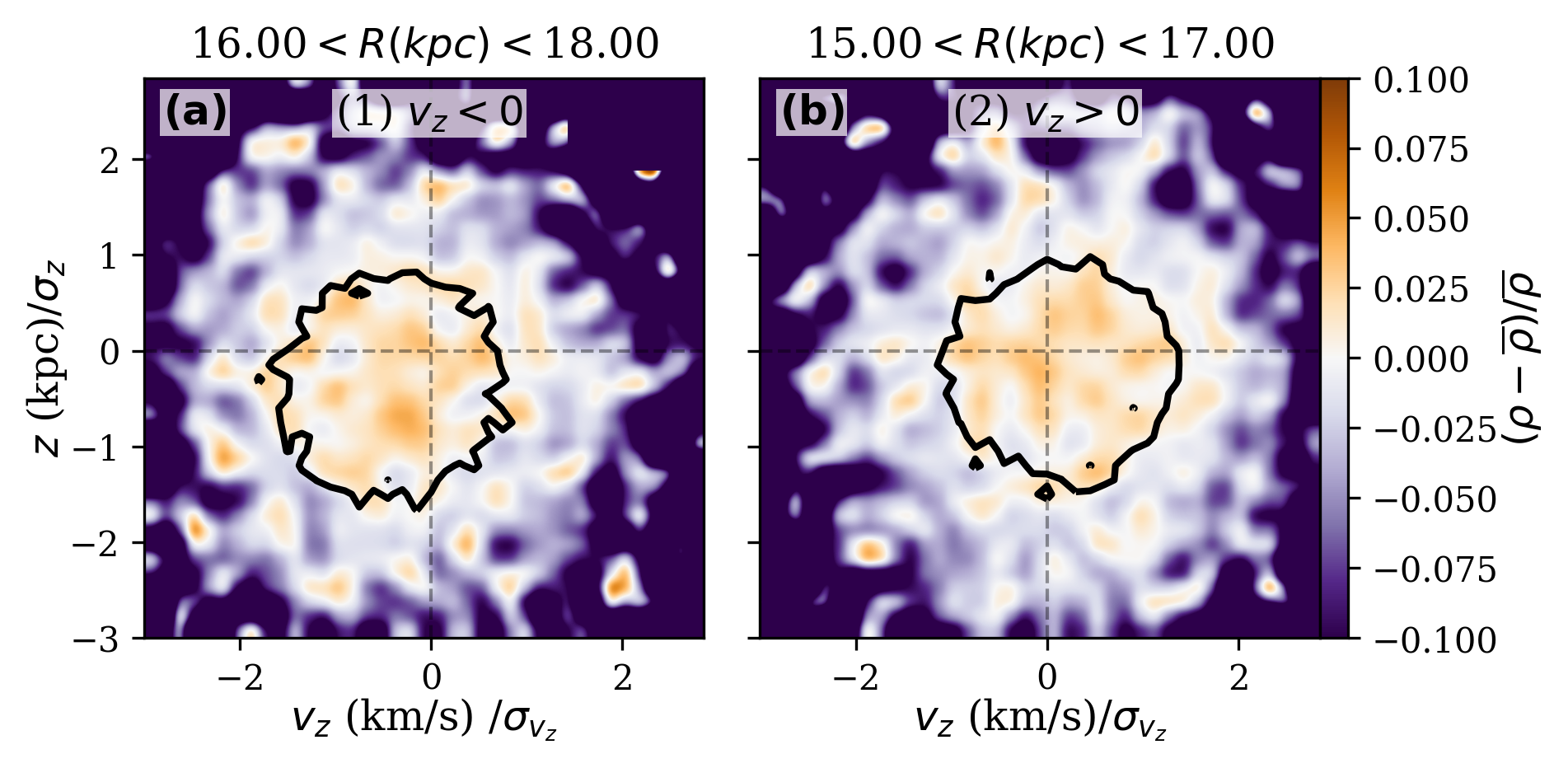}
    \caption{The regions with largest $|\Delvz|$ in simulations of the ongoing MW-Sgr interaction do not exhibit phase-spirals. \textbf{(a)} Corresponds to $v_z<0$ patch outlined in \figref{fig:xy_current_interaction_present_day_different_models}a. The black contour here encloses 50\% of the simulated stellar population in the patch. The sample is clearly offset toward $v_z<0$, and we also see that a significant fraction of stars in that region have $z<0$. \textbf{(b)} Corresponds to the $v_z>0$ patch in \figref{fig:xy_current_interaction_present_day_different_models}a. Once again, 50\% of the simulated sample is enclosed by the black contour here, which is clearly shifted toward $v_z>0$ (as expected) and $z<0$. This figure additionally provides $z$ information which is not apparent from \figref{fig:xy_current_interaction_present_day_different_models}a. }
    \label{fig: zvz_present_day_model_b}
\end{figure}

         

In Figures \ref{fig:xy_current_interaction_present_day_different_models}(a,b), we show a comparison of the \xy\ plane at present day colored by \vz$\times 10^{10}M_\odot/$\Msg~for \hyperref[tab:I]{Models B} and \hyperref[tab:I]{C}. There is a distinct pattern in \vz~which appears to be more or less consistent between the two panels, indicating that it is largely a result of the ongoing MW-Sgr interaction, and not a remnant of past disk crossings. The maximum \vz\ amplitude induced in the disk is significantly lower than the toy model ($|\max$ \vz$_{,toy}|\times 10^{10}\msun/\Msg\approx7/0.3$ \kmpers$ \approx 23$ \kmpers) because epicyclic oscillations average out the signal. Moreover, the amplitude of \vz$\propto M_{Sgr}$, such that maximum $|v_z|\times 10^{10}M_\odot/$\Msg$\approx3$ \kmpers\ remains almost the same in both panels.

Finally, \figref{fig:xy_current_interaction_present_day_different_models}c shows a self-consistent simulation snapshot (see \S\ref{subsec: bonsai sim description} for details of the model) at present day which has evolved over the past 6.88 Gyr, and the current Sgr remnant mass is $M_{Sgr, \text{rem}}\approx 8\tenexp{9}\msun$. Although the \xy\ extent of \figref{fig:xy_current_interaction_present_day_different_models}c is different from panels \ref{fig:xy_current_interaction_present_day_different_models}(a,b), in \textit{all three models}, a blue (\vz$<0$) patch is visible around $\phi\sim 0^{\circ}$ on the opposite side of the disk across from the Sun ($(x_\odot, y_\odot)=(-8,0)$), and a red ($v_z>0$) region emerges with the highest amplitude of positive \vz\ around $\phi\sim -90^\circ$. These two regions are roughly marked by black contours in panel \ref{fig:xy_current_interaction_present_day_different_models}a, and appear around the same azimuth in panels \ref{fig:xy_current_interaction_present_day_different_models}(b,c) as well (albeit in different \xy\ positions). The blue patch (labeled `(1)$v_z<0$' in panel \ref{fig:xy_current_interaction_present_day_different_models}a) is created by the downward pull of the Sgr-like satellite, it's present-day \xy\ location marked by a red dot in each panel. The red region (labeled `(2)$v_z>0$' in panel \ref{fig:xy_current_interaction_present_day_different_models}a) consists of simulated stars which were pulled downwards by Sgr $\sim 100$ Myrs ago, and are now traveling upwards as part of their vertical epicyclic motion.

The reason why the coordinates and \vz\ amplitude of the blue and red patches differ between the panels \ref{fig:xy_current_interaction_present_day_different_models}(a,b) versus panel \ref{fig:xy_current_interaction_present_day_different_models}c is that the disk potential and satellite orbit in the self-consistent simulation are different from those in the test particle simulation, and there is little control over these in the self-consistent model. We emphasize that the existence of the $v_z<0$ and $v_z>0$ regions not just in the test particle models but also in a self-consistent MW-Sgr interaction is main point of \figref{fig:xy_current_interaction_present_day_different_models}.  The fact that a self-consistent disk also exhibits a significant $v_z<0$ signal near the location of the dwarf is a strong indicator that this signature might exist in our own Galaxy. If detected in future surveys \citep[like the SDSS-V Milky Way Mapper;][]{sdssv_2019_mwm}, the amplitude could be used to infer \sag's remnant mass precisely while the shape could be used to infer trends in disk frequencies (and hence the force field) in that region. There is the caveat that the remnant mass ($\Msg \approx 8\times 10^{9}\msun$) inducing the \vz\ signal in \figref{fig:xy_current_interaction_present_day_different_models}c is much larger than the present day \Msg\ quoted in \citet{Vasiliev2020}. Thus it is possible that Sgr's tidal effects are negligible at present compared to the self-sustained bending waves in the disk.


\subsubsection{Phase-Spirals}
We do not find phase spirals in the test particle disk models due to the ongoing \sag-MW interaction, as might be anticipated since there is little time for these features to develop.
However, there is significant vertical asymmetry in the simulated disk caused by the interaction, which we present in \figref{fig: zvz_present_day_model_b}. Panels \ref{fig: zvz_present_day_model_b}(a,b) respectively show the \zvz\ plane for particles in the blue region (labeled `(1)$v_z<0$' in panel \ref{fig:xy_current_interaction_present_day_different_models}a) and the red region (labeled `(2)$v_z>0$' in panel \ref{fig:xy_current_interaction_present_day_different_models}a). The black contours in \ref{fig: zvz_present_day_model_b}(a,b) enclose 50\% of the simulated sample in each of the two regions. In panel \ref{fig: zvz_present_day_model_b}a, the sample is clearly offset toward $v_z<0$, and a significant fraction of simulated stars have $z<0$. Panel \ref{fig: zvz_present_day_model_b}b shows the black contour offset toward $v_z>0$ (as expected) and $z<0$, indicating that although stars in this patch have started to travel upward with positive \vz, they are still below the midplane.

\section{Conclusions}\label{sec:conc}
The main conclusions of this paper are the following:
\begin{enumerate}
    
    \item As shown in \citet{li2020}, selection of stars based on their azimuthal action \jphi\ (or equivalently, their guiding radius $R_g$) makes it possible to resolve multiple \zvz\ spiral morphologies (\figref{fig:jphi-covers-more}). We make the case that each of these local phase-spirals probes \textit{distinct regions} of the disk and began developing at \textit{distinct times} (\figref{fig:5-jphi-groups-zvz-plane-and-asym-param}). These multiple \zvz\ spirals can originate from the \textit{same perturbative event} (\figref{fig:5-jphi-groups-xy-plane-at-5-snaps}), and therefore  each of them offers a different perspective on the same event.
    \end{enumerate}
    
     Multiple \zvz\ spiral morphologies (over a wide range in \rg) exist in the local sample because the effects of a single satellite disk-crossing are long-lasting ($\sim300$ Myr) and affect the entire disk. The varied spirals are a reflection both of the fact that different regions of the disk experience the perturbation with different amplitudes (demonstrated with a toy model in \figref{fig:az-lag-lead-3stars}) and that different regions respond with different characteristic frequencies. Coincidence of the disk-crossing timescale and orbital time period can further amplify the distortions (\figsref{fig:freqs_and_time_periods},\ref{fig:asymmetry_parameters}).
     
    Since a single perturbative event can have such an extended impact on disk dynamics, we investigate the ongoing MW-\sag\ interaction. Our investigation leads to several insights about the current disk passage.
    \begin{enumerate}
        \item[2.]  Even though Sgr is not expected to cross the midplane for another $\sim 30$ Myr, the bulk of the influence on the inner disk ($\rgal \lesssim 10$  kpc) from this imminent passage has \textit{already happened} (\figref{fig: az_toy_model_Sgr_current_passage}).
        
        \item[3.] We do not find \zvz\ spirals in test particle simulations of the ongoing interaction (\figref{fig: zvz_present_day_model_b}), but significant asymmetry is expected in \rgal--$\phi$, leading to a disk-wide physical spiral (seen in \figref{fig:xy_current_interaction_present_day_different_models}). The amplitude of this \vz\ signature scales linearly with \Msg.
        
        \item[4.] The fact that there is a $\vz<0$ (blue) patch around Sgr's ``present-day"  \xy\ position, and a similarly sized $v_z>0$ (red) patch around $\phi\approx-90^{\circ}$ in \textit{both} the test particle and self-consistent disk (\figref{fig:xy_current_interaction_present_day_different_models}), suggests it is likely that these signatures are present in our Galaxy as well. These patches might be detectable in future Milky Way surveys \citep[e.g. SDSS-V Milky Way Mapper;][]{sdssv_2019_mwm}. If the true mass and orbit of Sgr are indeed such that there is a coherent \vz\ offset, the amplitude can be used to infer \Msg, while the morphology and location will help constrain both the properties of the disk and Sgr's orbit. However, it is possible that the projection shown in \figref{fig:xy_current_interaction_present_day_different_models}c is in reality obscured by pre-existing or independent disk dynamics.
    \end{enumerate}

We consider the above conclusions to be generic consequences of phase-mixing alone. A full interpretation of observed features will need to take account of the disk self-gravity, and signatures of the more recent interactions will need to be disentangled from prior perturbative events. Nevertheless, our results demonstrate intuitive starting points towards  building methods that can tease apart these  overlapping responses.

\section*{Acknowledgements}
The authors would like to thank {Chris Carr and Douglas Filho for their help}. SSG is funded by New York University through the MacCracken Fellowship. JASH and APW are supported by a Flatiron Research Fellowship at the Flatiron institute, which is supported by the Simons Foundation. This work was performed in part by JASH and APW at Aspen Center for Physics, which is supported by National Science Foundation grant PHY-1607611. KVJ was supported by NSF grant AST-1715582. CFPL acknowledges funding from the European Research Council (ERC) under the European Union's Horizon 2020 research and innovation program (grant agreement No. 852839). This work was supported in part by World Premier International Research Center Initiative (WPI Initiative), MEXT, Japan. This work was performed in part at Aspen Center for Physics, which is supported by National Science Foundation grant PHY-1607611. This work was partially supported by a grant from the Simons Foundation. 

\software{
    Astropy \citep{astropy, astropy:2018},
    galpy \citep{Bovy2015},
    gala \citep{gala, gala-v1_3},
    IPython \citep{ipython},
    matplotlib \citep{mpl},
    numpy \citep{numpy},
    scipy \citep{scipy}, pandas \citep[][]{pandas}.
}

\bibliography{main}{}
\bibliographystyle{aasjournal}

\end{document}